\documentclass[prb,aps,amsmath,amssymb,twocolumn]{revtex4-2}
\usepackage{graphicx,amsmath,amssymb,color,xcolor}
\usepackage{setspace}
\usepackage[utf8]{inputenc}
\usepackage{ulem}
\usepackage{nicefrac}
\usepackage{multirow, makecell, ragged2e}
\usepackage[titletoc,title]{appendix}
\usepackage[colorlinks,bookmarks=true,citecolor=blue,linkcolor=red,urlcolor=blue]{hyperref}
\usepackage{cancel}
\usepackage{bm}

\newcommand{\be}{\begin{equation}}
\newcommand{\ee}{\end{equation}}

\newcommand{\ba}{\begin{eqnarray}}
\newcommand{\ea}{\end{eqnarray}}

\newcommand{\LLL}{\text{P}_{\text{LLL}}}

\newcommand{\conj}[1]{#1{}^{\star}}

\def\beq{\begin{eqnarray}}
\def\eeq{\end{eqnarray}}

\newcommand{\abs}[1]{\vert #1 \vert}

\begin{document}
\title{Unlocking new regimes in fractional quantum Hall effect with quaternions}
\author{Mytraya Gattu and J. K. Jain}
\affiliation{Department of Physics, 104 Davey Lab, Pennsylvania State University, University Park, Pennsylvania 16802, USA}
\date{\today}
\begin{abstract}
We demonstrate that formulating the composite-fermion theory of the fractional quantum Hall (FQH) effect in terms of quaternions greatly expands its reach and opens the door into many interesting issues that were previously beyond the reach of quantitative theoretical investigation. As an illustration, we investigate the possibility of a nematic or a charge-density wave instability of the composite-fermion Fermi sea at half-filled Landau level and of the nearby FQH states by looking for a magneto-roton instability. Our quaternion formulation of the FQH effect has been inspired by mathematical developments in the theoretical analyses of gravitational wave modes and cosmic microwave background radiation, where an important role is played by spin-weighted spherical harmonics which are nothing but monopole harmonics appearing in the spherical geometry for the FQH effect. 
\end{abstract}
\maketitle

The fractional quantum Hall (FQH) effect~\cite{Tsui82} provides an example of a strongly correlated system for which an extremely accurate microscopic understanding has been achieved in terms of the composite fermions (CFs) which experience a reduced magnetic field~\cite{Jain89,Jain07,Halperin20}. The Jain CF wave functions for electrons at filling factor $\nu$ are constructed from the known wave functions of noninteracting electrons at filling factor $\nu^*$ [related by $\nu=\nu^*/(2p\nu^*\pm 1)$] by attaching $2p$ vortices followed by projection into the lowest Landau level (LLL). The LLL projection was initially carried out by a brute force expansion of the unprojected wave functions in the Fock basis~\cite{Dev92, Wu93}. However, this could be accomplished for very small systems because the dimension of the Fock basis grows (super) exponentially with the number of electrons ($N$) and soon becomes larger than what can be stored on a computer. For the same reason, exact diagonalization (ED) studies are also typically restricted to systems with $N\leq 16$ electrons. These allowed theoretical study of FQH states at $\nu=1/3, 2/3, 2/5, 3/5$ [because, it requires larger and larger systems to construct states with larger $n$ along $\nu=n/(2n+1)$.]

In a significant leap, Jain and Kamilla (JK)~\cite{Jain97,Jain97b} introduced an ansatz for the LLL projected wave function which does not require expansion into Fock space basis functions and allows study of systems with up to 100-200 CFs over a large range of filling factors. Extensive tests against ED studies showed these wave functions to be remarkably accurate~\cite{Jain97,Jain97b,Balram13,Jain20}. The JK approach enabled detailed comparisons between theory and experiment for numerous quantities which include: energy gaps~\cite{Du93,Manoharan94,Rosales21,Zhao22}, energies of neutral magneto-roton modes~\cite{Pinczuk93,Scarola00,Kang01,Dujovne05,Moller05,Kukushkin09,Rhone11,Balram22,Majumder14}, spin or valley polarization transitions~\cite{Du95,Du97,Park98,Kukushkin99,Bishop07,Padmanabhan10,Davenport12,Liu15,Zhang16}, competition between FQH liquid and crystal as a function of the filling factor and Landau level (LL) mixing~\cite{Archer13,Liu14a,Zhao18,Rosales21b,Madathil24}, Halperin-Lee-Read (HLR) Fermi sea at $\nu=1/2$~\cite{Halperin93,Rezayi94,Rezayi00,Lee18,Fremling18,Geraedts18},
pairing instability of the HLR Fermi sea and even denominator FQH effect (FQHE)~\cite{Sharma24,Henderson23,Moller08,Moran12,Suen94b,Balram18,Wagner21,Wang22,Wang23,Yutushui24}, phase diagram of bilayer FQHE~\cite{Li19,Liu19,Scarola01b,Moller08,Moller09,Zaletel18,Li17,Wagner21,Wagner24,Hu24}, scanning tunneling microscopy of FQH states~\cite{Hu23,Jain05,Gattu23,Pu24a,Yue24}, entanglement~\cite{Rodriguez12,Davenport15,Shao15}, etc. The JK approach also has limitations: its computational accuracy and speed diminish as we go to larger $n$ along the Jain $\nu=n/(2n+1)$ states. As a result, accurate predictions from this method are restricted in practice to states in the range $-5 \leq n \leq 7$.

We report in this Letter the next leap, which has been inspired by parallel developments in the fields of gravitational waves and cosmic microwave background radiation, where a generalization of the spherical harmonics called spin-weighted spherical harmonics (SWSHs)\cite{Newman66, Thorne80} play an essential role. Of particular relevance here is Boyle's work, wherein he showed that representing the SWSHs in terms of quaternions greatly simplifies the theoretical analysis~\cite{Boyle16}. Though not widely appreciated, the SWSHs are nothing but the monopole harmonics~\cite{Wu76,Wu77, Dray85} of Haldane's spherical geometry~\cite{Haldane83} used in the studies of the FQHE. We show that a quaternion formulation of the Jain FQH states suggests a strategy for accurate and efficient evaluation that dramatically enlarges the parameter space where  reliable results may be obtained within the CF theory.  With the quaternion formulation, we are able to investigate the Jain $\nu = n/(2n + 1)$ FQH states for at least $-20 \leq n \leq 20$ for systems containing more than $400$ electrons. We show below results for $N=390$ particles at $\nu=15/31$, whose Fock basis has a dimension of $\sim 10^{236}$.

\begin{figure}
    \includegraphics[width=0.48\columnwidth]{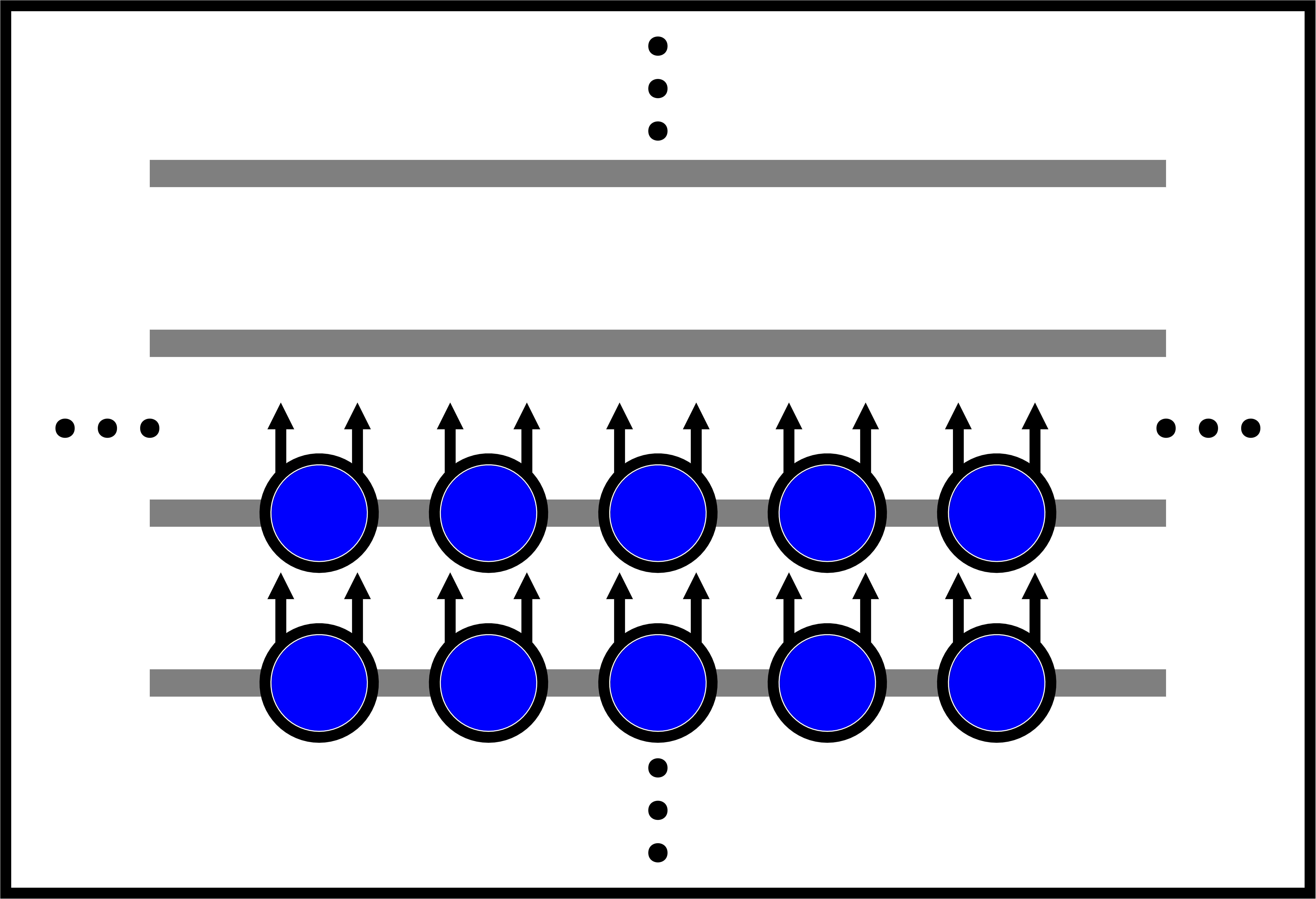}
    \includegraphics[width=0.48\columnwidth]{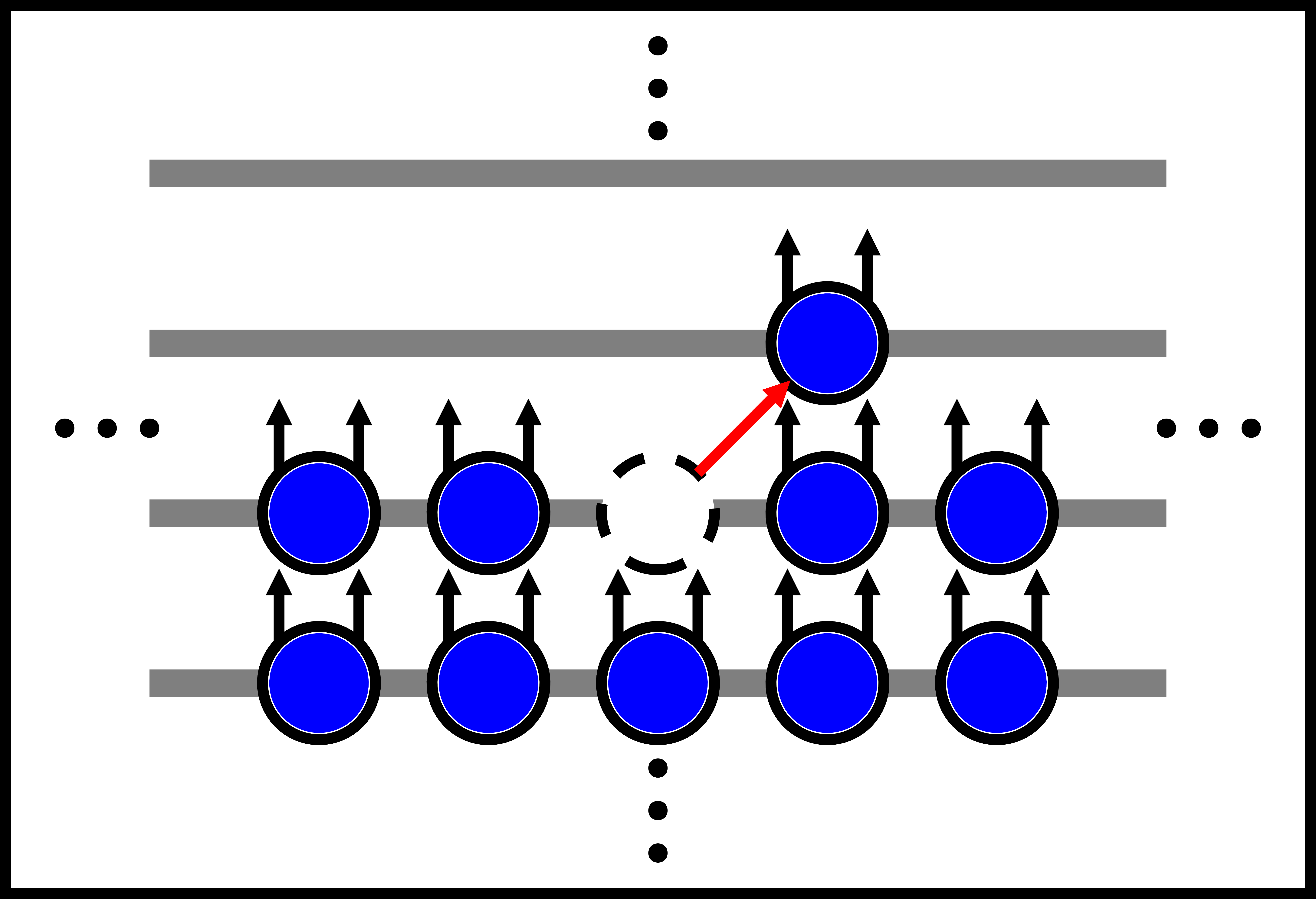}
    \caption{The schematic on the left illustrates an incompressible state of composite fermions (CFs), depicted as electrons decorated with two flux quanta, at filling factor $\nu = n/(2n+1)$, containing $n$ fully occupied CF LLs. The schematic on the right depicts the CF exciton at $\nu = n/(2n+1)$, which is the lowest energy neutral excitation.}
    \label{fig:cf-gs-exciton}
\end{figure}

As an illustration of our new method, we address whether the FQHE at $\nu=n/(2n\pm 1)$ at large $n$ or the HLR Fermi sea at $\nu=1/2$ is unstable to a charge density wave or a nematic phase~\cite{Maciejko13, Nguyen18, Yang20, You14, Yang20, Pu24}. For this purpose, we evaluate the dispersions of the neutral magneto-roton modes, modeled as CF excitons~\cite{Dev92,Scarola00} shown schematically in Fig.~\ref{fig:cf-gs-exciton}, for states up to $\nu=15/31$. Our study brings out interesting features in the  dispersion
and demonstrates an extraordinary robustness of the FQHE along $n/(2n+1)$, in agreement with experiments that reveal more and more Jain fractions as the sample quality is enhanced~\cite{Chung21,Chung22,Wang23}. Furthermore, our study does not find any evidence for instability of the HLR Fermi sea at $\nu=1/2$.

In Haldane's spherical geometry, electrons are confined to move on the surface $\mathcal{S}$ of a sphere, with a magnetic monopole of strength $Q$ located at its center. This monopole generates a uniform radial magnetic field of magnitude $B = 2Q\phi_0 /(4\pi R^2)$ and produces a magnetic flux of $2Q \phi_0$ through the surface $\mathcal{S}$, where $\phi_0=hc/e$ is called the flux quantum. The single-particle wave functions in this geometry are the monopole harmonics 
\begin{equation} \label{eq:monopole-harmonics-sphere}
\begin{split}
 &Y_{Q,l,m}(\theta, \phi) = N_{Q,l,m}(-1)^{l-m}v^{Q-m}u^{Q+m}\\
 &\times \sum_{s}(-1)^{s}\tbinom{l-Q}{s}\tbinom{l+Q}{l-m-s}(\conj{v}v)^{l-Q-s}(\conj{u}u)^{s}\\
  &N_{Q, l, m} = \sqrt{[(2l+1)/4\pi]\times [\tbinom{2l}{l+Q}/\tbinom{2l}{l+m}]}
 \end{split}
\end{equation}
Here $l = \abs{Q} + n_{l}$ is the angular momentum, $m = -l, -l+1, \dots, l$ is the azimuthal quantum number i.e. the $L_{z}$ eigenvalue, $n_l=0, 1, \cdots$ is the LL index, $\Omega = (\theta, \phi)$ denotes the electron’s position in spherical coordinates, and $u = \cos (\theta/2) e^{\imath \phi / 2}$ and $v = \sin (\theta / 2) e^{-\imath \phi / 2}$ are the spinor variables.

The monopole harmonics $Y_{Q, l, m}$ furnish a projective representation of the three-dimensional rotational symmetry on $\mathcal{S}$. Under an active rotation $\mathcal{R}$ defined by the Euler angles $\alpha, \beta, \gamma$ (about the $z-y-z$ axes), the monopole harmonics at $(\theta, \phi)$ can be related to the monopole harmonics at $(\theta^{\prime}, \phi^{\prime})$ (where $(\theta, \phi)$ goes to $(\theta^{\prime}, \phi^{\prime})$ under the rotation) as follows~\cite{Wu76, Wu77, Boyle16} 
\begin{equation}\label{eq:monopole-harmonics-sphere-rotation}
    \begin{split}
        &Y_{Q, l, m}(\theta^{\prime}, \phi^{\prime})\\
        &= e^{-\imath Q \psi^{\prime}}\sum_{m^{\prime}=-l}^{l}(-1)^{m-m^{\prime}}[D_{m,m^{\prime}}^{l}(\alpha, \beta, \gamma)]^{\star}Y_{Q, l, m^{\prime}}(\theta, \phi)      
    \end{split}
\end{equation}
Here, $D^{l}_{m, m^{\prime}}(\alpha, \beta, \gamma)$ is the unitary Wigner-$D$ matrix corresponding to $\mathcal{R}$~\cite{Wigner31, Sakurai20}. The phase $\psi^{\prime}$ is a function of both $\mathcal{R}$ and $(\theta, \phi)$~\cite{Wu77, Boyle16}; although it may be explicitly evaluated, its discontinuous nature makes it challenging to exploit the rotational symmetry on $\mathcal{S}$ in both analytical and numerical calculations, as discussed in the Suplementary Materials (SM)~\cite{SM-Gattu-2024}. To overcome this, Boyle~\cite{Boyle16} proposed an ingenious re-definition of the monopole harmonics as functions of unit quaternions that transform ordinarily under rotations and more generally under quaternion multiplication.

We begin by recalling the essential properties of quaternions. A quaternion $r \equiv (r_0, r_1, r_2, r_3)$, where $r_0, r_1, r_2, r_3 \in \mathbb{R}$, is expressed as $r = r_0\bm{1} + r_1\bm{i} + r_2\bm{j} + r_3\bm{k}$. The basis elements $\bm{1}, \bm{i}, \bm{j}, \bm{k}$ correspond to the Pauli matrices $\sigma_0, \imath\sigma_3, \imath\sigma_2, \imath\sigma_1$. The rules for quaternion addition and multiplication follow from those of the basis elements, i.e., the Pauli matrices. The inverse of a (non-zero) quaternion is defined in terms of its conjugate $r^{\star} \equiv (r_{0}, -r_{1}, -r_{2}, -r_{3})$ and norm $||r|| \equiv \sqrt{r^{\star} \cdot r} = \sqrt{r_{0}^{2}+r_{1}^{2}+r_{2}^{2}+r_{3}^{2}}$ as $r^{-1} = r^{\star} / ||r||^{2}$.

We now specialize to the set of unit quaternions given by $\left\{r \vert (r_{0}, r_{1}, r_{2}, r_{3}) \in \mathbb{R}^{4} \ni r_{0}^{2} + r_{1}^{2} + r_{2}^{2} + r_{3}^{2} = 1\right\}$ which is isomorphic to $\mathbb{S}^{3}$. We can parameterize $r$ as
\begin{equation}\label{eq:quaternion-sphere-mapping}
    \begin{split}
        r &= e^{\phi \bm{k}/2}e^{\theta \bm{j}/2}e^{\psi \bm{k}/2}\\
        &= \cos{\frac \theta 2} \cos \frac{\phi+\psi}{2}\bm{1} - \sin{\frac \theta 2} \sin \frac{\phi-\psi}{2}\bm{i} \\
        &+ \sin{\frac \theta 2} \cos \frac{\phi-\psi}{2}\bm{j} + \cos{\frac \theta 2} \sin \frac{\phi + \psi}{2}\bm{k} \\
    \end{split}
\end{equation}
where $(\theta, \phi, \psi)$ are the three angular coordinates on $\mathbb{S}^{3}$. 

We now define, following Boyle, 
\begin{equation}
    \mathcal{Y}_{Q,l,m}(r)=Y_{Q, l, m}(u\rightarrow r_{\mathrm{S}}, v\rightarrow r_{\mathrm{A}})
\end{equation}
where $r_{\mathrm{S}}$ and $r_{\mathrm{A}}$, the symmetric and anti-symmetric projections of $r$, are defined as 
\begin{equation}\label{eq:quaternion-projections}
    \begin{split}
        r_{\mathrm{S}} &= (r + \mathbf{k}\cdot r \cdot \mathbf{k}^{-1})/2 = r_{0} + r_{3}\mathbf{k}\\
        & = \cos(\theta/2)e^{\mathbf{k} (\phi + \psi) / 2}\\
        r_{\mathrm{A}} &= (r - \mathbf{k} \cdot r\cdot \mathbf{k}^{-1})/2 = r_{2} + r_{1}\mathbf{k}\\
        & = \sin(\theta/2)e^{\mathbf{k} (\psi-\phi)/2}
    \end{split}.
\end{equation}
(Within the sub-algebra generated by $\bm{1}$ and $\bm{k}$, $\bm{k}$ behaves exactly like $\imath$ i.e.~$\left[\bm{1}, \bm{k}\right]=0$ and $\bm{k}^{2}=-\bm{1}$. We will use $\bm{k}$ and $\imath$ interchangeably whenever the context is clear.)
It is also useful to define another function $\mathcal{D}^{l}_{m, m^{\prime}}(r) = \sqrt{4\pi/(2l+1)}\mathcal{Y}_{m^{\prime}, l, m}(r)$ that extends the Wigner-$D$ matrices to the space of unit quaternions. We can verify from Eq.~\ref{eq:monopole-harmonics-sphere}, that
\begin{equation} \label{eq:quaternion-sphere-monopole-harmonics-relation}
\begin{split}
        &\mathcal{Y}_{Q, l, m}(r)=e^{\imath Q \psi}Y_{Q,l,m}(\theta, \phi)\\
        &\mathcal{D}^{l}_{m, m^{\prime}}(r) = D^{l}_{m^{\prime}, m}(-\psi, \theta, -\phi).
\end{split}
\end{equation}
From the transformation of the projections $r_{\mathrm{S}}$ and $r_{\mathrm{A}}$ under multiplication,
\begin{equation}\label{eq:sym-anti-sym-quaternion}
    (r \cdot s)_{S} = r_{S}s_{S}-\conj{r}_{A}s_{A}, \;(r \cdot s)_{A} = r_{A}s_{S}+\conj{r}_{S}s_{A},
\end{equation}
it can be shown that both $\mathcal{D}^{l}_{m, m^{\prime}}(r)$ and $\mathcal{Y}_{Q,l,m}(r)$ obey the following transformation rule
\begin{equation}\label{eq:monopole-harmonics-rotation}
\begin{split}
    &\mathcal{D}^{l}_{m, m^{\prime}}(r) = \sum_{m^{\prime \prime}=-l}^{l}\mathcal{D}^{l}_{m, m^{\prime \prime}}(r \cdot s^{-1})\mathcal{D}^{l}_{m^{\prime \prime}, m^{\prime}}(s) \\
    &\mathcal{Y}_{Q, l, m}(r) = \sum_{m^{\prime}=-l}^{l}\mathcal{D}^{l}_{m, m^{\prime}}(r\cdot s^{-1})\mathcal{Y}_{Q, l, m^{\prime}}(s)\\
\end{split},
\end{equation}
which contains no phase factor! That is, $\mathcal{Y}_{Q, l, m}(r)$ transform ordinarily under (unit) quaternion multiplication (which would be referred to as rotation below, as it connects two points on $\mathbb{S}^{3}$). This simple behavior under rotation will play a crucial role below.

We now show that the quaternion formulation dramatically enlarges the range of FQH states that are accessible to quantitative study within CF theory. Let us recall the basic principle of the CF theory. It first prepares a state of $N$ electrons at an effective flux $2Q^*=2Q-2(N-1)$, which in the simplest cases is given by a Slater determinant $\det [Y_{Q^{\star}, l_{j}, m_{j}}(\Omega_{i})]$, where $\Omega_i$ is the position of the $i^{\rm th}$ electron. We then multiply it by $\prod_{j<k=1}^{N}(u_{j}v_{k}-u_{k}v_{j})^{2}$, which attaches $2$ vortices to each electron to convert it into a CF. Finally, this product must be projected into the LLL, which is conveniently accomplished \`a la the JK projection, which makes the ansatz:
\begin{equation}\label{eq:jk-projection-definition}
\begin{split}
    &\LLL \left\{\det{\left[Y_{Q^{\star}, l_{j}, m_{j}}(\Omega_{i})\right]}\prod_{j<k=1}^{N}(u_{j}v_{k}-u_{k}v_{j})^{2}\right\}\\
    &=\det \left[ \LLL Y_{Q^{\star}, l_{j}, m_{j}}(\Omega_{i})J_i\right]
\end{split}
\end{equation} 
where $\LLL$ is the LLL projection operator and $J_{i} = \prod_{k \neq i}(u_{i}v_{k}-u_{k}v_{i})$. In other words, instead of projecting the full wave function, we separately project each element of a determinant. Note that in $Y_{Q^{\star}, l_{j}, m_{j}}(\Omega_{i})J_i$ only one (the $i^{\rm th}$) electron can be outside the LLL, thereby simplifying LLL projection. 

Traditionally, the JK projection is carried out as~\cite{Jain97, Jain07},
\begin{equation}\label{eq:jk-projection-traditional}
    \begin{split}
    &\LLL Y_{Q^{\star}, l, m}(\Omega_{i})J_{i} \propto \sum_{s=0}(-1)^{s}\tbinom{l-Q^{\star}}{s}\times \\
    &\tbinom{Q^{\star}+l}{l-m-s}u_{i}^{Q^{\star}+m+s}v_{i}^{l-m-s}\partial^{s}_{u_{i}}\partial^{l-Q^{\star}-s}_{v_{i}}J_{i}
    \end{split}
\end{equation}
Substituting in Eq.~\ref{eq:jk-projection-definition} yields the CF Slater determinant state.  A limiting factor of this approach is the computation of $(l_{\mathrm{max}}-Q^{\star}+1)(l_{\mathrm{max}}-Q^{\star}+2)/2$ number of mixed derivatives [where $l_{\mathrm{max}}$ is the maximum angular momentum of an electron in $\det{Y_{Q^{\star}},l_{j},m_{j}}(\Omega_{i})$], which, combined with the numerical instability in their computation~\cite{Davenport12}, renders this approach unsuitable for systems with $n\geq 8$ (i.e. $l_{\mathrm{max}}-Q^{\star}=8$) or $\nu \geq 8/17$ (see SM~\cite{SM-Gattu-2024} for more details). We will now see how this limitation can be overcome by a new approach to JK projection which naturally follows from the quaternion extension of JK projected wave functions.

Remembering that any physical wave function $\Psi(\Omega_{1}, \dots, \Omega_{N})$ can be written in terms of monopole harmonics at a given $Q$, Eq.~\ref{eq:quaternion-sphere-monopole-harmonics-relation} implies that the quaternion extension of $\Psi$ can be written as $\Psi(\Omega_{1}, \dots, \Omega_{N})=\exp[-\imath Q (\sum_i\psi_{i})] \Psi(r_{1}, \dots, r_{N})$, where $\Psi(r_{1}, \dots, r_{N})$ (here, $r_{i} = e^{\phi_{i}\bm{k}/2}e^{\theta_{i}\bm{j}/2}e^{\psi_{i}\bm{k}/2}$) is obtained from $\Psi(\Omega_{1}, \dots, \Omega_{N})$ by replacing $Y_{Q, l, m}(\Omega_{i})$ with $\mathcal{Y}_{Q, l, m}(r_{i})$. In other words, Eq.~\ref{eq:quaternion-sphere-monopole-harmonics-relation} implies that the extra degrees of freedom $\psi_{i}$, introduced when we extend our wave functions from $\mathbb{S}^{2}$ to $\mathbb{S}^{3}$, \textit{do not contribute to any physical observable}. What we gain by extending to quaternions—beyond the simplification of transformations under rotations—is the insight that the ubiquitous Jastrow factor $u_iv_j-v_iu_j$ which maps into $d(r_{i}, r_{j}) = r_{i}{}_{\mathrm{S}}r_{j}{}_{\mathrm{A}} - r_{i}{}_{\mathrm{A}}r_{j}{}_{\mathrm{S}} = (r^{-1}_{i}\cdot r_{j})_{\mathrm{A}}$, can be regarded as a quaternion displacement that is invariant under (left) quaternion multiplication! We can exploit this invariance to evaluate $\LLL \mathcal{Y}_{Q^{\star}, l, m}(r_{i})\prod_{j \neq i}d(r_{i}, r_{j})$ which appears in the quaternion extension of Eq.~\ref{eq:jk-projection-definition}, as $\sum_{m^{\prime}}\mathcal{D}^{l}_{m, m^{\prime}}(r^{-1})\LLL \mathcal{Y}_{Q^{\star}, l, m^{\prime}}(r \cdot r_{i})\prod_{j \neq i}d(r \cdot r_{i}, r \cdot r_{j})$ for arbitrary $r$. This allows us to rotate the system to bring the $i^{\rm th}$ electron to a convenient point, perform LLL projection, and rotate back. The crucial observation underlying our new approach is that the LLL projection is especially simple and efficient at $r_i=\bm 1$ which naturally suggests the choice $r = r_{i}^{-1}$ and
allows us to write the elements in the determinant of the JK projected wave functions as (see SM~\cite{SM-Gattu-2024} for a detailed derivation):
\begin{equation}\label{eq:jk-projection-new}
    \begin{split}
    &\LLL \mathcal{Y}_{Q^{\star}, l, m}(r_{i})\prod_{j \neq i}d(r_{i}, r_{j})\\
    &= \sum_{m^{\prime}=Q^{\star}}^{l}(-1)^{m^{\prime}-Q^{\star}}N^{l}_{m^{\prime}, Q^{\star}, Q_{1}}\mathcal{D}^{l}_{m, m^{\prime}}(r_{i})\\
    &\times \tilde{e}_{m^{\prime}-Q^{\star}}(\left\{{(r_{i}^{-1}\cdot r_{j})_{\mathrm{S}}}/{(r_{i}^{-1}\cdot r_{j})_{\mathrm{A}}}\right\}_{j \neq i})\prod_{j \neq i}d(r_{i}, r_{j})\\
    &N^{l}_{m^{\prime}, Q^{\star}, Q_{1}} = \frac{\tbinom{2Q_{1}}{l-Q^{\star}}\tbinom{l-Q^{\star}}{m^{\prime}-Q^{\star}}}{\tbinom{2Q_{1}+l+Q^{\star}+1}{l-Q^{\star}}}\sqrt{\frac{(2l+1)}{4\pi}\frac{\tbinom{2l}{l+Q^{\star}}}{\tbinom{2l}{l+m^{\prime}}}}
    \end{split}.
\end{equation}
Here, $Q_{1} = (N-1)/2$ and $\tilde{e}_{m^{\prime}-Q^{\star}}$ is the elementary symmetric polynomial~\cite{Macdonald79} of degree $m^{\prime}-Q^{\star}$ in the $N-1$ variables $\left\{(r_{i}^{-1}\cdot r_{j})_{\mathrm{S}}/(r_{i}^{-1}\cdot r_{j})_{\mathrm{A}}\right\}_{j \neq i}$ divided by $\tbinom{N-1}{m^{\prime}-Q^{\star}}$. The key advantage of our new approach is that the computation of all the required elementary symmetric polynomials which is the rate limiting step, scales only as $l_{\mathrm{max}}-Q^{\star}$ and is robust to numerical instability~\cite{Rehman11}. This significantly increases both the computational speed and accuracy (further enhanced by the method proposed in Ref.~\cite{Feng15}) of the JK projected wave functions, making it possible to study FQH systems containing hundreds of electrons between fillings $1/3 \leq \nu \leq 20/41$ and $20/39 \leq \nu \leq 2/3$ (see SM~\cite{SM-Gattu-2024} for more details).
\begin{figure}
    \includegraphics[width=\columnwidth]{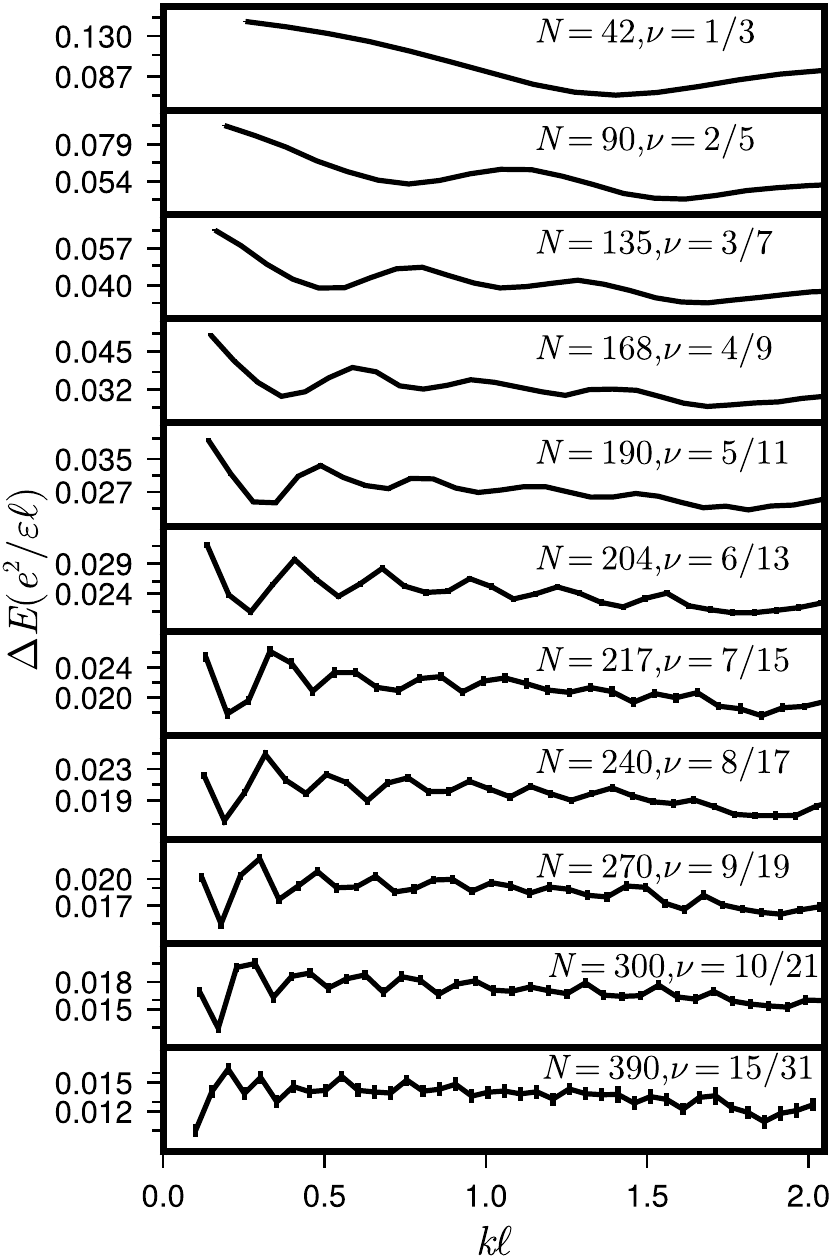}
    \caption{The plots show the 
    excitation energies $\Delta E$ of the lowest neutral magneto-roton modes as a function of the wave vector $k$ for the Jain FQH states at filling factors $\nu = n/(2n+1)$, obtained using our new method. The system size $N$ and the filling factor $\nu$ are shown on each panel. The energies are for a two-dimensional system, given in units of $e^{2}/\varepsilon \ell$, where $\ell = \sqrt{\hbar c/eB}$ is the magnetic length. The Monte Carlo uncertainty in $\Delta E$ is indicated by vertical bars.
    }
    \label{fig:collected-dispersion}
\end{figure}

As a first application of our method, we look for a magneto-roton instability of the FQHE at $\nu=n/(2n+1)$ as we approach $\nu=1/2$. While the CF theory predicts the possibility of FQHE at all fractions of the form $\nu=n/(2pn\pm 1)$, detailed calculations and comparisons with experiments are required to ascertain whether the Coulomb interaction stabilizes these states. [The Coulomb interaction fails to stabilize the $n/(2pn\pm1)$ states for very large $2p$ in the LLL or for any $2p$ in high LLs.] The FQHE along $\nu=n/(2n+1)$ would survive to arbitrary high $n$ if the CFs were completely non-interacting, but the CFs do have a residual interaction between them, which can, in principle, cause an instability when the interaction strength is comparable to the gap.

Remarkably, through its explicit wave functions, the CF theory also provides an accurate account of the residual CF-CF interaction and can, in principle, reveal an instability. We ask here if the FQHE along $\nu=n/(2n+1)$ is destabilized by the softening of the neutral  magneto-roton excitation\cite{Girvin85},
accurately described as a CF exciton (a CF particle-hole pair)~\cite{Dev92,Scarola00,Balram24}. Fig.~\ref{fig:collected-dispersion} shows the magneto-roton dispersions for several fractions, including for $\nu=15/31$. (We have not made here a serious effort to achieve the largest $n$ or $N$.) These dispersions show a rich structure appearing in the form of $n$ approximately equally-spaced minima~(except at $\nu = 15/31$, where larger systems would be needed to identify all minima), as predicted by field theoretical treatments of CFs~\cite{Simon93, Golkar16, Park00,Balram17c}, but no instability is encountered. Note that the first minimum occurs at $k\ell\rightarrow 0$~\cite{Balram24}, and the last at $k\ell\rightarrow 2$, which is equivalent to $k= 2k_{\rm F}$ at $\nu=1/2$ (where $k_{\rm F}=1/\ell$).

\begin{figure}
    \includegraphics[width=0.49\columnwidth]{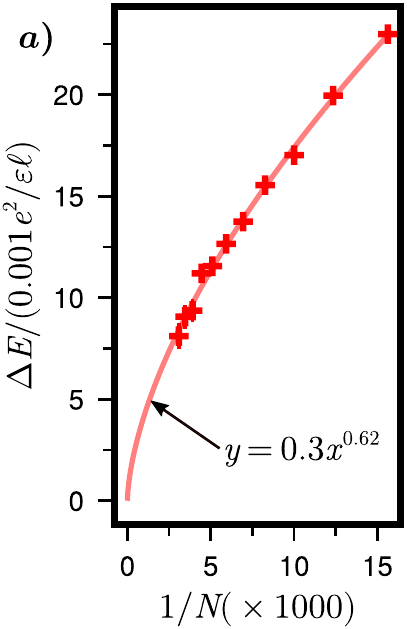}
    \includegraphics[width=0.49\columnwidth]{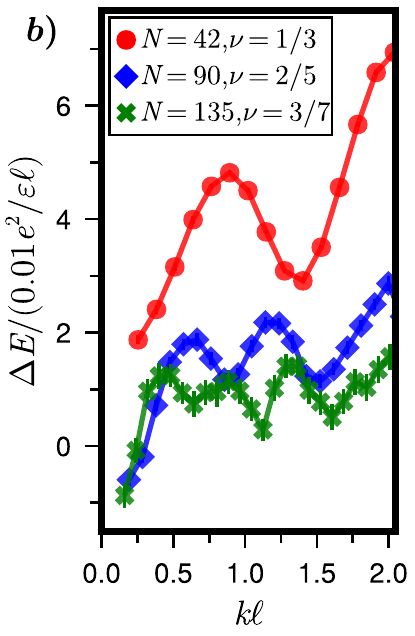}
    \caption{
    a) Excitation energy $\Delta E$ of the $k \rightarrow 0$ CF exciton for filled shell states of CFs at $Q^*=0$. The state with $n$ filled shells has $N=n^2$ CFs. For all systems considered here ($n=8, 9, \dots, 18 \Leftrightarrow N=64, 81, \dots, 324$), the $k \rightarrow 0$ CF exciton is also the lowest energy excitation. b) Excitation energy $\Delta E$ of CF excitons as a function of $k \ell$ for Jain FQH states at $\nu = 1/3, 2/5, 3/7$ in the second LL.
    }
    \label{fig:graviton-mode-qstar-0}
\end{figure}

If the magneto-roton energy were to become negative for $\nu=n/(2n+1)$ at some large $n$, that would imply a similar behavior for the CF Fermi sea at $\nu=1/2$.
The CF Fermi sea is obtained for CFs at $Q^*=0$ in the limit $N\rightarrow \infty$. We consider systems with $n$ filled angular momentum shells with $N=n^2$ particles. For $n\geq 8$, the $k\ell\rightarrow 0$ magneto-roton has the lowest energy; should its energy become negative, that would indicate a nematic instability~\cite{Maciejko13, Yang20, You14, Pu24}.
Fig.~\ref{fig:graviton-mode-qstar-0}a shows that its energy approaches zero (rather than a negative value) in the limit of $N\rightarrow \infty$, and thus does not indicate any instability of the CF Fermi sea. These calculations demonstrate that FQHE remains surprisingly robust along the Jain sequence $\nu=n/(2n+1)$. 
As a demonstration of the non-triviality of these results, we plot in Fig.~\ref{fig:graviton-mode-qstar-0}b the CF exciton dispersion at certain fractions in the second LL (of GaAs system), where we use the effective interaction given in Ref.~\cite{Yutushui24} to mimic the second LL. Here, an instability is seen at $\nu=2/5, 3/7$ etc.  We note that the Jain wave functions are not very accurate representations of the exact states in the second LL, and thus our conclusions are less reliable than those in the LLL. Nonetheless, the instability of the Jain 2/5 state in the second LL is consistent with studies that point to the Read-Rezayi state here~\cite{Read99,Bonderson12,Sreejith13,Repellin15,Kusmierz18,Balram22a}. No FQHE has been observed at $3/7,\; 4/9,\;\cdots$ in the second LL.

We thank Ajit Balram and Steven Kivelson for illuminating discussions. 
Computations for this research were performed on the Pennsylvania State University's Institute for Computational and Data Sciences' Roar Collab supercomputer. Plots were made using Makie~\cite{Danisch21}.
\bibliographystyle{unsrtnat}
\bibliography{biblio_fqhe.bib}

\begin{thebibliography}{102}
\providecommand{\natexlab}[1]{#1}
\providecommand{\url}[1]{\texttt{#1}}
\expandafter\ifx\csname urlstyle\endcsname\relax
  \providecommand{\doi}[1]{doi: #1}\else
  \providecommand{\doi}{doi: \begingroup \urlstyle{rm}\Url}\fi

\bibitem[Tsui et~al.(1982)Tsui, Stormer, and Gossard]{Tsui82}
D.~C. Tsui, H.~L. Stormer, and A.~C. Gossard.
\newblock Two-dimensional magnetotransport in the extreme quantum limit.
\newblock \emph{Phys. Rev. Lett.}, 48:\penalty0 1559--1562, May 1982.
\newblock \doi{10.1103/PhysRevLett.48.1559}.
\newblock URL \url{http://link.aps.org/doi/10.1103/PhysRevLett.48.1559}.

\bibitem[Jain(1989)]{Jain89}
J.~K. Jain.
\newblock Composite-fermion approach for the fractional quantum {Hall} effect.
\newblock \emph{Phys. Rev. Lett.}, 63:\penalty0 199--202, Jul 1989.
\newblock \doi{10.1103/PhysRevLett.63.199}.
\newblock URL \url{http://link.aps.org/doi/10.1103/PhysRevLett.63.199}.

\bibitem[Jain(2007)]{Jain07}
J.~K. Jain.
\newblock \emph{Composite Fermions}.
\newblock Cambridge University Press, New York, US, 2007.

\bibitem[Halperin and Jain(2020)]{Halperin20}
Bertrand~I Halperin and Jainendra~K Jain, editors.
\newblock \emph{{Fractional} {Quantum} {Hall} {Effects} {New} {Developments}}.
\newblock World Scientific, 2020.
\newblock \doi{10.1142/11751}.
\newblock URL \url{https://worldscientific.com/doi/abs/10.1142/11751}.

\bibitem[Dev and Jain(1992)]{Dev92}
Gautam Dev and J.~K. Jain.
\newblock Band structure of the fractional quantum {Hall} effect.
\newblock \emph{Phys. Rev. Lett.}, 69:\penalty0 2843--2846, Nov 1992.
\newblock \doi{10.1103/PhysRevLett.69.2843}.
\newblock URL \url{http://link.aps.org/doi/10.1103/PhysRevLett.69.2843}.

\bibitem[Wu et~al.(1993)Wu, Dev, and Jain]{Wu93}
X.~G. Wu, G.~Dev, and J.~K. Jain.
\newblock Mixed-spin incompressible states in the fractional quantum hall
  effect.
\newblock \emph{Physical Review Letters}, 71\penalty0 (1):\penalty0 153–156,
  July 1993.
\newblock ISSN 0031-9007.
\newblock \doi{10.1103/physrevlett.71.153}.
\newblock URL \url{http://dx.doi.org/10.1103/PhysRevLett.71.153}.

\bibitem[Jain and Kamilla(1997{\natexlab{a}})]{Jain97}
J.~K. Jain and R.~K. Kamilla.
\newblock Composite fermions in the {Hilbert} space of the lowest electronic
  {Landau} level.
\newblock \emph{Int. J. Mod. Phys. B}, 11\penalty0 (22):\penalty0 2621--2660,
  1997{\natexlab{a}}.
\newblock \doi{10.1142/S0217979297001301}.

\bibitem[Jain and Kamilla(1997{\natexlab{b}})]{Jain97b}
J.~K. Jain and R.~K. Kamilla.
\newblock Quantitative study of large composite-fermion systems.
\newblock \emph{Phys. Rev. B}, 55:\penalty0 R4895--R4898, Feb
  1997{\natexlab{b}}.
\newblock \doi{10.1103/PhysRevB.55.R4895}.
\newblock URL \url{http://link.aps.org/doi/10.1103/PhysRevB.55.R4895}.

\bibitem[Balram et~al.(2013)Balram, W\'ojs, and Jain]{Balram13}
Ajit~C. Balram, Arkadiusz W\'ojs, and Jainendra~K. Jain.
\newblock State counting for excited bands of the fractional quantum {Hall}
  effect: Exclusion rules for bound excitons.
\newblock \emph{Phys. Rev. B}, 88:\penalty0 205312, Nov 2013.
\newblock \doi{10.1103/PhysRevB.88.205312}.
\newblock URL \url{http://link.aps.org/doi/10.1103/PhysRevB.88.205312}.

\bibitem[Jain(2020)]{Jain20}
J.~K. Jain.
\newblock Thirty {Years} of {Composite} {Fermions} and {Beyond}.
\newblock In B.~I. Halperin and J.~K. Jain, editors, \emph{Fractional Quantum
  Hall Effects: New Developments}, chapter~1, pages 1--78. World Scientific Pub
  Co Inc, Singapore, 2020.
\newblock \doi{10.1142/9789811217494_0001}.
\newblock URL
  \url{https://www.worldscientific.com/doi/abs/10.1142/9789811217494_0001}.

\bibitem[Du et~al.(1993)Du, Stormer, Tsui, Pfeiffer, and West]{Du93}
R.~R. Du, H.~L. Stormer, D.~C. Tsui, L.~N. Pfeiffer, and K.~W. West.
\newblock Experimental evidence for new particles in the fractional quantum
  {Hall} effect.
\newblock \emph{Phys. Rev. Lett.}, 70:\penalty0 2944--2947, May 1993.
\newblock \doi{10.1103/PhysRevLett.70.2944}.
\newblock URL \url{http://link.aps.org/doi/10.1103/PhysRevLett.70.2944}.

\bibitem[Manoharan et~al.(1994)Manoharan, Shayegan, and Klepper]{Manoharan94}
H.~C. Manoharan, M.~Shayegan, and S.~J. Klepper.
\newblock Signatures of a novel {Fermi} liquid in a two-dimensional composite
  particle metal.
\newblock \emph{Phys. Rev. Lett.}, 73:\penalty0 3270--3273, Dec 1994.
\newblock \doi{10.1103/PhysRevLett.73.3270}.
\newblock URL \url{http://link.aps.org/doi/10.1103/PhysRevLett.73.3270}.

\bibitem[Villegas~Rosales et~al.(2021{\natexlab{a}})Villegas~Rosales, Madathil,
  Chung, Pfeiffer, West, Baldwin, and Shayegan]{Rosales21}
K.~A. Villegas~Rosales, P.~T. Madathil, Y.~J. Chung, L.~N. Pfeiffer, K.~W.
  West, K.~W. Baldwin, and M.~Shayegan.
\newblock Fractional quantum hall effect energy gaps: Role of electron layer
  thickness.
\newblock \emph{Phys. Rev. Lett.}, 127:\penalty0 056801, Jul
  2021{\natexlab{a}}.
\newblock \doi{10.1103/PhysRevLett.127.056801}.
\newblock URL \url{https://link.aps.org/doi/10.1103/PhysRevLett.127.056801}.

\bibitem[Zhao et~al.(2022)Zhao, Kudo, Faugno, Balram, and Jain]{Zhao22}
Tongzhou Zhao, Koji Kudo, W.~N. Faugno, Ajit~C. Balram, and J.~K. Jain.
\newblock Revisiting excitation gaps in the fractional quantum hall effect.
\newblock \emph{Phys. Rev. B}, 105:\penalty0 205147, May 2022.
\newblock \doi{10.1103/PhysRevB.105.205147}.
\newblock URL \url{https://link.aps.org/doi/10.1103/PhysRevB.105.205147}.

\bibitem[Pinczuk et~al.(1993)Pinczuk, Dennis, Pfeiffer, and West]{Pinczuk93}
A.~Pinczuk, B.~S. Dennis, L.~N. Pfeiffer, and K.~West.
\newblock Observation of collective excitations in the fractional quantum
  {Hall} effect.
\newblock \emph{Phys. Rev. Lett.}, 70:\penalty0 3983--3986, Jun 1993.
\newblock \doi{10.1103/PhysRevLett.70.3983}.
\newblock URL \url{http://link.aps.org/doi/10.1103/PhysRevLett.70.3983}.

\bibitem[Scarola et~al.(2000)Scarola, Park, and Jain]{Scarola00}
Vito~W. Scarola, Kwon Park, and Jainendra~K. Jain.
\newblock Rotons of composite fermions: Comparison between theory and
  experiment.
\newblock \emph{Phys. Rev. B}, 61:\penalty0 13064--13072, May 2000.
\newblock \doi{10.1103/PhysRevB.61.13064}.
\newblock URL \url{http://link.aps.org/doi/10.1103/PhysRevB.61.13064}.

\bibitem[Kang et~al.(2001)Kang, Pinczuk, Dennis, Pfeiffer, and West]{Kang01}
Moonsoo Kang, A.~Pinczuk, B.~S. Dennis, L.~N. Pfeiffer, and K.~W. West.
\newblock Observation of multiple magnetorotons in the fractional quantum
  {Hall} effect.
\newblock \emph{Phys. Rev. Lett.}, 86:\penalty0 2637--2640, Mar 2001.
\newblock \doi{10.1103/PhysRevLett.86.2637}.
\newblock URL \url{http://link.aps.org/doi/10.1103/PhysRevLett.86.2637}.

\bibitem[Dujovne et~al.(2005)Dujovne, Pinczuk, Kang, Dennis, Pfeiffer, and
  West]{Dujovne05}
Irene Dujovne, A.~Pinczuk, Moonsoo Kang, B.~S. Dennis, L.~N. Pfeiffer, and
  K.~W. West.
\newblock Composite-fermion spin excitations as $\ensuremath{\nu}$ approaches
  $1/2$: Interactions in the {Fermi} sea.
\newblock \emph{Phys. Rev. Lett.}, 95:\penalty0 056808, Jul 2005.
\newblock \doi{10.1103/PhysRevLett.95.056808}.
\newblock URL \url{http://link.aps.org/doi/10.1103/PhysRevLett.95.056808}.

\bibitem[M\"oller and Simon(2005)]{Moller05}
Gunnar M\"oller and Steven~H. Simon.
\newblock Composite fermions in a negative effective magnetic field: A {Monte}
  {Carlo} study.
\newblock \emph{Phys. Rev. B}, 72:\penalty0 045344, Jul 2005.
\newblock \doi{10.1103/PhysRevB.72.045344}.
\newblock URL \url{http://link.aps.org/doi/10.1103/PhysRevB.72.045344}.

\bibitem[Kukushkin et~al.(2009)Kukushkin, Smet, Scarola, Umansky, and von
  Klitzing]{Kukushkin09}
Igor~V. Kukushkin, Jurgen~H. Smet, Vito~W. Scarola, Vladimir Umansky, and Klaus
  von Klitzing.
\newblock Dispersion of the excitations of fractional quantum {Hall} states.
\newblock \emph{Science}, 324\penalty0 (5930):\penalty0 1044--1047, 2009.
\newblock \doi{10.1126/science.1171472}.
\newblock URL \url{http://www.sciencemag.org/content/324/5930/1044.abstract}.

\bibitem[Rhone et~al.(2011)Rhone, Majumder, Dennis, Hirjibehedin, Dujovne,
  Groshaus, Gallais, Jain, Mandal, Pinczuk, Pfeiffer, and West]{Rhone11}
Trevor~D. Rhone, Dwipesh Majumder, Brian~S. Dennis, Cyrus Hirjibehedin, Irene
  Dujovne, Javier~G. Groshaus, Yann Gallais, Jainendra~K. Jain, Sudhansu~S.
  Mandal, Aron Pinczuk, Loren Pfeiffer, and Ken West.
\newblock Higher-energy composite fermion levels in the fractional quantum
  {Hall} effect.
\newblock \emph{Phys. Rev. Lett.}, 106:\penalty0 096803, Mar 2011.
\newblock \doi{10.1103/PhysRevLett.106.096803}.
\newblock URL \url{http://link.aps.org/doi/10.1103/PhysRevLett.106.096803}.

\bibitem[Balram et~al.(2022)Balram, Liu, Gromov, and
  Papi\ifmmode~\acute{c}\else \'{c}\fi{}]{Balram22}
Ajit~C. Balram, Zhao Liu, Andrey Gromov, and Zlatko Papi\ifmmode~\acute{c}\else
  \'{c}\fi{}.
\newblock Very-high-energy collective states of partons in fractional quantum
  hall liquids.
\newblock \emph{Phys. Rev. X}, 12:\penalty0 021008, Apr 2022.
\newblock \doi{10.1103/PhysRevX.12.021008}.
\newblock URL \url{https://link.aps.org/doi/10.1103/PhysRevX.12.021008}.

\bibitem[Majumder and Mandal(2014)]{Majumder14}
Dwipesh Majumder and Sudhansu~S. Mandal.
\newblock Neutral collective modes in spin-polarized fractional quantum {Hall}
  states at filling factors $\frac{1}{3}$, $\frac{2}{5}$, $\frac{3}{7}$, and
  $\frac{4}{9}$.
\newblock \emph{Phys. Rev. B}, 90:\penalty0 155310, Oct 2014.
\newblock \doi{10.1103/PhysRevB.90.155310}.
\newblock URL \url{http://link.aps.org/doi/10.1103/PhysRevB.90.155310}.

\bibitem[Du et~al.(1995)Du, Yeh, Stormer, Tsui, Pfeiffer, and West]{Du95}
R.~R. Du, A.~S. Yeh, H.~L. Stormer, D.~C. Tsui, L.~N. Pfeiffer, and K.~W. West.
\newblock Fractional quantum {Hall} effect around $\nu=3/2$: Composite fermions
  with a spin.
\newblock \emph{Phys. Rev. Lett.}, 75:\penalty0 3926--3929, Nov 1995.
\newblock \doi{10.1103/PhysRevLett.75.3926}.
\newblock URL \url{http://link.aps.org/doi/10.1103/PhysRevLett.75.3926}.

\bibitem[Du et~al.(1997)Du, Yeh, Stormer, Tsui, Pfeiffer, and West]{Du97}
R.~R. Du, A.~S. Yeh, H.~L. Stormer, D.~C. Tsui, L.~N. Pfeiffer, and K.~W. West.
\newblock g factor of composite fermions around $\nu=3/2$ from
  angular-dependent activation-energy measurements.
\newblock \emph{Phys. Rev. B}, 55:\penalty0 R7351--R7354, Mar 1997.
\newblock \doi{10.1103/PhysRevB.55.R7351}.
\newblock URL \url{http://link.aps.org/doi/10.1103/PhysRevB.55.R7351}.

\bibitem[Park and Jain(1998)]{Park98}
K.~Park and J.~K. Jain.
\newblock Phase diagram of the spin polarization of composite fermions and a
  new effective mass.
\newblock \emph{Phys. Rev. Lett.}, 80:\penalty0 4237--4240, May 1998.
\newblock \doi{10.1103/PhysRevLett.80.4237}.
\newblock URL \url{http://link.aps.org/doi/10.1103/PhysRevLett.80.4237}.

\bibitem[Kukushkin et~al.(1999)Kukushkin, v.~Klitzing, and Eberl]{Kukushkin99}
I.~V. Kukushkin, K.~v.~Klitzing, and K.~Eberl.
\newblock Spin polarization of composite fermions: Measurements of the {Fermi}
  energy.
\newblock \emph{Phys. Rev. Lett.}, 82:\penalty0 3665--3668, May 1999.
\newblock \doi{10.1103/PhysRevLett.82.3665}.
\newblock URL \url{http://link.aps.org/doi/10.1103/PhysRevLett.82.3665}.

\bibitem[Bishop et~al.(2007)Bishop, Padmanabhan, Vakili, Shkolnikov,
  De~Poortere, and Shayegan]{Bishop07}
N.~C. Bishop, M.~Padmanabhan, K.~Vakili, Y.~P. Shkolnikov, E.~P. De~Poortere,
  and M.~Shayegan.
\newblock Valley polarization and susceptibility of composite fermions around a
  filling factor $\nu=3/2$.
\newblock \emph{Phys. Rev. Lett.}, 98:\penalty0 266404, Jun 2007.
\newblock \doi{10.1103/PhysRevLett.98.266404}.
\newblock URL \url{http://link.aps.org/doi/10.1103/PhysRevLett.98.266404}.

\bibitem[Padmanabhan et~al.(2010)Padmanabhan, Gokmen, and
  Shayegan]{Padmanabhan10}
Medini Padmanabhan, T.~Gokmen, and M.~Shayegan.
\newblock Composite fermion valley polarization energies: Evidence for
  particle-hole asymmetry.
\newblock \emph{Phys. Rev. B}, 81:\penalty0 113301, Mar 2010.
\newblock \doi{10.1103/PhysRevB.81.113301}.
\newblock URL \url{http://link.aps.org/doi/10.1103/PhysRevB.81.113301}.

\bibitem[Davenport and Simon(2012)]{Davenport12}
Simon~C. Davenport and Steven~H. Simon.
\newblock Spinful composite fermions in a negative effective field.
\newblock \emph{Phys. Rev. B}, 85:\penalty0 245303, Jun 2012.
\newblock \doi{10.1103/PhysRevB.85.245303}.
\newblock URL \url{http://link.aps.org/doi/10.1103/PhysRevB.85.245303}.

\bibitem[Liu et~al.(2015)Liu, Hasdemir, Shabani, Shayegan, Pfeiffer, West, and
  Baldwin]{Liu15}
Yang Liu, S.~Hasdemir, J.~Shabani, M.~Shayegan, L.~N. Pfeiffer, K.~W. West, and
  K.~W. Baldwin.
\newblock Multicomponent fractional quantum {Hall} states with subband and spin
  degrees of freedom.
\newblock \emph{Phys. Rev. B}, 92:\penalty0 201101, Nov 2015.
\newblock \doi{10.1103/PhysRevB.92.201101}.
\newblock URL \url{http://link.aps.org/doi/10.1103/PhysRevB.92.201101}.

\bibitem[Zhang et~al.(2016)Zhang, Wójs, and Jain]{Zhang16}
Yuhe Zhang, A.~Wójs, and J.~K. Jain.
\newblock Landau-level mixing and particle-hole symmetry breaking for spin
  transitions in the fractional quantum hall effect.
\newblock \emph{Physical Review Letters}, 117\penalty0 (11), September 2016.
\newblock ISSN 1079-7114.
\newblock \doi{10.1103/physrevlett.117.116803}.
\newblock URL \url{http://dx.doi.org/10.1103/PhysRevLett.117.116803}.

\bibitem[Archer et~al.(2013)Archer, Park, and Jain]{Archer13}
Alexander~C. Archer, Kwon Park, and Jainendra~K. Jain.
\newblock Competing crystal phases in the lowest {Landau} level.
\newblock \emph{Phys. Rev. Lett.}, 111:\penalty0 146804, Oct 2013.
\newblock \doi{10.1103/PhysRevLett.111.146804}.

\bibitem[Liu et~al.(2014)Liu, Kamburov, Hasdemir, Shayegan, Pfeiffer, West, and
  Baldwin]{Liu14a}
Yang Liu, D.~Kamburov, S.~Hasdemir, M.~Shayegan, L.~N. Pfeiffer, K.~W. West,
  and K.~W. Baldwin.
\newblock Fractional quantum {Hall} effect and {Wigner} crystal of interacting
  composite fermions.
\newblock \emph{Phys. Rev. Lett.}, 113:\penalty0 246803, Dec 2014.
\newblock \doi{10.1103/PhysRevLett.113.246803}.
\newblock URL \url{http://link.aps.org/doi/10.1103/PhysRevLett.113.246803}.

\bibitem[Zhao et~al.(2018)Zhao, Zhang, and Jain]{Zhao18}
Jianyun Zhao, Yuhe Zhang, and J.~K. Jain.
\newblock Crystallization in the fractional quantum hall regime induced by
  landau-level mixing.
\newblock \emph{Physical Review Letters}, 121\penalty0 (11), September 2018.
\newblock ISSN 1079-7114.
\newblock \doi{10.1103/physrevlett.121.116802}.
\newblock URL \url{http://dx.doi.org/10.1103/PhysRevLett.121.116802}.

\bibitem[Villegas~Rosales et~al.(2021{\natexlab{b}})Villegas~Rosales, Singh,
  Ma, Hossain, Chung, Pfeiffer, West, Baldwin, and Shayegan]{Rosales21b}
K.~A. Villegas~Rosales, S.~K. Singh, Meng~K. Ma, Md.~Shafayat Hossain, Y.~J.
  Chung, L.~N. Pfeiffer, K.~W. West, K.~W. Baldwin, and M.~Shayegan.
\newblock Competition between fractional quantum hall liquid and wigner solid
  at small fillings: Role of layer thickness and landau level mixing.
\newblock \emph{Phys. Rev. Res.}, 3:\penalty0 013181, Feb 2021{\natexlab{b}}.
\newblock \doi{10.1103/PhysRevResearch.3.013181}.
\newblock URL \url{https://link.aps.org/doi/10.1103/PhysRevResearch.3.013181}.

\bibitem[Madathil et~al.(2024)Madathil, Wang, Singh, Gupta, Rosales, Chung,
  West, Baldwin, Pfeiffer, Engel, and Shayegan]{Madathil24}
P.~T. Madathil, C.~Wang, S.~K. Singh, A.~Gupta, K.~A.~Villegas Rosales, Y.~J.
  Chung, K.~W. West, K.~W. Baldwin, L.~N. Pfeiffer, L.~W. Engel, and
  M.~Shayegan.
\newblock Signatures of correlated defects in an ultraclean wigner crystal in
  the extreme quantum limit.
\newblock \emph{Phys. Rev. Lett.}, 132:\penalty0 096502, Mar 2024.
\newblock \doi{10.1103/PhysRevLett.132.096502}.
\newblock URL \url{https://link.aps.org/doi/10.1103/PhysRevLett.132.096502}.

\bibitem[Halperin et~al.(1993)Halperin, Lee, and Read]{Halperin93}
B.~I. Halperin, Patrick~A. Lee, and Nicholas Read.
\newblock Theory of the half-filled {Landau} level.
\newblock \emph{Phys. Rev. B}, 47:\penalty0 7312--7343, Mar 1993.
\newblock \doi{10.1103/PhysRevB.47.7312}.
\newblock URL \url{http://link.aps.org/doi/10.1103/PhysRevB.47.7312}.

\bibitem[Rezayi and Read(1994)]{Rezayi94}
E.~Rezayi and N.~Read.
\newblock Fermi-liquid-like state in a half-filled {Landau} level.
\newblock \emph{Phys. Rev. Lett.}, 72:\penalty0 900--903, Feb 1994.
\newblock \doi{10.1103/PhysRevLett.72.900}.
\newblock URL \url{http://link.aps.org/doi/10.1103/PhysRevLett.72.900}.

\bibitem[Rezayi and Haldane(2000)]{Rezayi00}
E.~H. Rezayi and F.~D.~M. Haldane.
\newblock Incompressible paired {Hall} state, stripe order, and the composite
  fermion liquid phase in half-filled {Landau} levels.
\newblock \emph{Phys. Rev. Lett.}, 84:\penalty0 4685--4688, May 2000.
\newblock \doi{10.1103/PhysRevLett.84.4685}.
\newblock URL \url{http://link.aps.org/doi/10.1103/PhysRevLett.84.4685}.

\bibitem[Lee et~al.(2018)Lee, Shao, Kim, Haldane, and Rezayi]{Lee18}
Kyungmin Lee, Junping Shao, Eun-Ah Kim, F.~D.~M. Haldane, and Edward~H. Rezayi.
\newblock Pomeranchuk instability of composite fermi liquids.
\newblock \emph{Phys. Rev. Lett.}, 121:\penalty0 147601, Oct 2018.
\newblock \doi{10.1103/PhysRevLett.121.147601}.
\newblock URL \url{https://link.aps.org/doi/10.1103/PhysRevLett.121.147601}.

\bibitem[Fremling et~al.(2018)Fremling, Moran, Slingerland, and
  Simon]{Fremling18}
M.~Fremling, N.~Moran, J.~K. Slingerland, and S.~H. Simon.
\newblock Trial wave functions for a composite {Fermi} liquid on a torus.
\newblock \emph{Phys. Rev. B}, 97:\penalty0 035149, Jan 2018.
\newblock \doi{10.1103/PhysRevB.97.035149}.
\newblock URL \url{https://link.aps.org/doi/10.1103/PhysRevB.97.035149}.

\bibitem[Geraedts et~al.(2018)Geraedts, Wang, Rezayi, and Haldane]{Geraedts18}
Scott~D. Geraedts, Jie Wang, E.~H. Rezayi, and F.~D.~M. Haldane.
\newblock Berry phase and model wave function in the half-filled {Landau}
  level.
\newblock \emph{Phys. Rev. Lett.}, 121:\penalty0 147202, Oct 2018.
\newblock \doi{10.1103/PhysRevLett.121.147202}.
\newblock URL \url{https://link.aps.org/doi/10.1103/PhysRevLett.121.147202}.

\bibitem[Sharma et~al.(2024)Sharma, Balram, and Jain]{Sharma24}
Anirban Sharma, Ajit~C. Balram, and J.~K. Jain.
\newblock Composite-fermion pairing at half-filled and quarter-filled lowest
  landau level.
\newblock \emph{Phys. Rev. B}, 109:\penalty0 035306, Jan 2024.
\newblock \doi{10.1103/PhysRevB.109.035306}.
\newblock URL \url{https://link.aps.org/doi/10.1103/PhysRevB.109.035306}.

\bibitem[Henderson et~al.(2023)Henderson, M\"oller, and Simon]{Henderson23}
Greg~J. Henderson, Gunnar M\"oller, and Steven~H. Simon.
\newblock Energy minimization of paired composite fermion wave functions in the
  spherical geometry.
\newblock \emph{Phys. Rev. B}, 108:\penalty0 245128, Dec 2023.
\newblock \doi{10.1103/PhysRevB.108.245128}.
\newblock URL \url{https://link.aps.org/doi/10.1103/PhysRevB.108.245128}.

\bibitem[M\"oller and Simon(2008)]{Moller08}
G.~M\"oller and S.~H. Simon.
\newblock Paired composite-fermion wave functions.
\newblock \emph{Phys. Rev. B}, 77:\penalty0 075319, Feb 2008.
\newblock \doi{10.1103/PhysRevB.77.075319}.
\newblock URL \url{https://link.aps.org/doi/10.1103/PhysRevB.77.075319}.

\bibitem[Moran et~al.(2012)Moran, Sterdyniak, Vidanovi\ifmmode~\acute{c}\else
  \'{c}\fi{}, Regnault, and Milovanovi\ifmmode~\acute{c}\else
  \'{c}\fi{}]{Moran12}
N.~Moran, A.~Sterdyniak, I.~Vidanovi\ifmmode~\acute{c}\else \'{c}\fi{},
  N.~Regnault, and M.~V. Milovanovi\ifmmode~\acute{c}\else \'{c}\fi{}.
\newblock Topological $d$-wave pairing structures in jain states.
\newblock \emph{Phys. Rev. B}, 85:\penalty0 245307, Jun 2012.
\newblock \doi{10.1103/PhysRevB.85.245307}.
\newblock URL \url{http://link.aps.org/doi/10.1103/PhysRevB.85.245307}.

\bibitem[Suen et~al.(1994)Suen, Manoharan, Ying, Santos, and Shayegan]{Suen94b}
Y.~W. Suen, H.~C. Manoharan, X.~Ying, M.~B. Santos, and M.~Shayegan.
\newblock Origin of the $\nu=1/2$ fractional quantum {Hall} state in wide
  single quantum wells.
\newblock \emph{Phys. Rev. Lett.}, 72:\penalty0 3405--3408, May 1994.
\newblock \doi{10.1103/PhysRevLett.72.3405}.
\newblock URL \url{https://link.aps.org/doi/10.1103/PhysRevLett.72.3405}.

\bibitem[Balram et~al.(2018)Balram, Barkeshli, and Rudner]{Balram18}
Ajit~C. Balram, Maissam Barkeshli, and Mark~S. Rudner.
\newblock Parton construction of a wave function in the anti-{Pfaffian} phase.
\newblock \emph{Phys. Rev. B}, 98:\penalty0 035127, Jul 2018.
\newblock \doi{10.1103/PhysRevB.98.035127}.
\newblock URL \url{https://link.aps.org/doi/10.1103/PhysRevB.98.035127}.

\bibitem[Wagner et~al.(2021)Wagner, Nguyen, Simon, and Halperin]{Wagner21}
Glenn Wagner, Dung~X. Nguyen, Steven~H. Simon, and Bertrand~I. Halperin.
\newblock $s$-wave paired electron and hole composite fermion trial state for
  quantum hall bilayers with $\ensuremath{\nu}=1$.
\newblock \emph{Phys. Rev. Lett.}, 127:\penalty0 246803, Dec 2021.
\newblock \doi{10.1103/PhysRevLett.127.246803}.
\newblock URL \url{https://link.aps.org/doi/10.1103/PhysRevLett.127.246803}.

\bibitem[Wang et~al.(2022)Wang, Gupta, Singh, Chung, Pfeiffer, West, Baldwin,
  Winkler, and Shayegan]{Wang22}
Chengyu Wang, A.~Gupta, S.~K. Singh, Y.~J. Chung, L.~N. Pfeiffer, K.~W. West,
  K.~W. Baldwin, R.~Winkler, and M.~Shayegan.
\newblock Even-denominator fractional quantum hall state at filling factor
  $\ensuremath{\nu}=3/4$.
\newblock \emph{Phys. Rev. Lett.}, 129:\penalty0 156801, Oct 2022.
\newblock \doi{10.1103/PhysRevLett.129.156801}.
\newblock URL \url{https://link.aps.org/doi/10.1103/PhysRevLett.129.156801}.

\bibitem[Wang et~al.(2023)Wang, Gupta, Singh, Madathil, Chung, Pfeiffer,
  Baldwin, Winkler, and Shayegan]{Wang23}
Chengyu Wang, A.~Gupta, S.~K. Singh, P.~T. Madathil, Y.~J. Chung, L.~N.
  Pfeiffer, K.~W. Baldwin, R.~Winkler, and M.~Shayegan.
\newblock Fractional quantum hall state at filling factor
  $\ensuremath{\nu}=1/4$ in ultra-high-quality gaas two-dimensional hole
  systems.
\newblock \emph{Phys. Rev. Lett.}, 131:\penalty0 266502, Dec 2023.
\newblock \doi{10.1103/PhysRevLett.131.266502}.
\newblock URL \url{https://link.aps.org/doi/10.1103/PhysRevLett.131.266502}.

\bibitem[Yutushui and Mross(2024)]{Yutushui24}
Misha Yutushui and David~F Mross.
\newblock Phase diagram of compressible and paired states in the quarter-filled
  landau level.
\newblock \emph{arXiv [cond-mat.str-el]}, August 2024.
\newblock URL \url{http://dx.doi.org/10.1103/PhysRevB.92.235302}.

\bibitem[Li et~al.(2019)Li, Shi, Zeng, Watanabe, Taniguchi, Hone, and
  Dean]{Li19}
J.~I.~A. Li, Q.~Shi, Y.~Zeng, K.~Watanabe, T.~Taniguchi, J.~Hone, and C.~R.
  Dean.
\newblock Pairing states of composite fermions in double-layer graphene.
\newblock \emph{Nature Physics}, 15\penalty0 (9):\penalty0 898--903, 2019.
\newblock ISSN 1745-2481.
\newblock \doi{10.1038/s41567-019-0547-z}.
\newblock URL \url{https://doi.org/10.1038/s41567-019-0547-z}.

\bibitem[Liu et~al.(2019)Liu, Hao, Watanabe, Taniguchi, Halperin, and
  Kim]{Liu19}
Xiaomeng Liu, Zeyu Hao, Kenji Watanabe, Takashi Taniguchi, Bertrand~I.
  Halperin, and Philip Kim.
\newblock Interlayer fractional quantum {Hall} effect in a coupled graphene
  double layer.
\newblock \emph{Nature Physics}, 15\penalty0 (9):\penalty0 893--897, 2019.
\newblock ISSN 1745-2481.
\newblock \doi{10.1038/s41567-019-0546-0}.
\newblock URL \url{https://doi.org/10.1038/s41567-019-0546-0}.

\bibitem[Scarola and Jain(2001)]{Scarola01b}
V.~W. Scarola and J.~K. Jain.
\newblock Phase diagram of bilayer composite fermion states.
\newblock \emph{Phys. Rev. B}, 64:\penalty0 085313, Aug 2001.
\newblock \doi{10.1103/PhysRevB.64.085313}.
\newblock URL \url{http://link.aps.org/doi/10.1103/PhysRevB.64.085313}.

\bibitem[M\"oller et~al.(2009)M\"oller, Simon, and Rezayi]{Moller09}
Gunnar M\"oller, Steven~H. Simon, and Edward~H. Rezayi.
\newblock Trial wave functions for $\ensuremath{\nu}=\frac{1}{2}+\frac{1}{2}$
  quantum hall bilayers.
\newblock \emph{Phys. Rev. B}, 79:\penalty0 125106, Mar 2009.
\newblock \doi{10.1103/PhysRevB.79.125106}.
\newblock URL \url{https://link.aps.org/doi/10.1103/PhysRevB.79.125106}.

\bibitem[Zaletel et~al.(2018)Zaletel, Geraedts, Papić, and Rezayi]{Zaletel18}
Michael~P. Zaletel, Scott Geraedts, Zlatko Papić, and Edward~H. Rezayi.
\newblock Evidence for a topological “exciton fermi sea” in bilayer
  graphene.
\newblock \emph{Physical Review B}, 98\penalty0 (4), July 2018.
\newblock ISSN 2469-9969.
\newblock \doi{10.1103/physrevb.98.045113}.
\newblock URL \url{http://dx.doi.org/10.1103/PhysRevB.98.045113}.

\bibitem[Li et~al.(2017)Li, Tan, Chen, Zeng, Taniguchi, Watanabe, Hone, and
  Dean]{Li17}
J.~I.~A. Li, C.~Tan, S.~Chen, Y.~Zeng, T.~Taniguchi, K.~Watanabe, J.~Hone, and
  C.~R. Dean.
\newblock Even denominator fractional quantum {Hall} states in bilayer
  graphene.
\newblock \emph{Science}, 2017.
\newblock ISSN 0036-8075.
\newblock \doi{10.1126/science.aao2521}.
\newblock URL
  \url{http://science.sciencemag.org/content/early/2017/10/04/science.aao2521}.

\bibitem[Wagner and Nguyen(2024)]{Wagner24}
Glenn Wagner and Dung~X. Nguyen.
\newblock Successive electron-vortex binding in quantum hall bilayers at
  $\ensuremath{\nu}=\frac{1}{4}+\frac{3}{4}$.
\newblock \emph{Phys. Rev. B}, 110:\penalty0 195106, Nov 2024.
\newblock \doi{10.1103/PhysRevB.110.195106}.
\newblock URL \url{https://link.aps.org/doi/10.1103/PhysRevB.110.195106}.

\bibitem[Hu et~al.(2024)Hu, Neupert, and Wagner]{Hu24}
Qi~Hu, Titus Neupert, and Glenn Wagner.
\newblock Single-parameter variational wave functions for quantum hall
  bilayers.
\newblock \emph{Phys. Rev. B}, 109:\penalty0 155138, Apr 2024.
\newblock \doi{10.1103/PhysRevB.109.155138}.
\newblock URL \url{https://link.aps.org/doi/10.1103/PhysRevB.109.155138}.

\bibitem[Hu et~al.(2023)Hu, Tsui, He, Kamber, Wang, Mohammadi, Watanabe,
  Taniguchi, Papic, Zaletel, and Yazdani]{Hu23}
Yuwen Hu, Yen-Chen Tsui, Minhao He, Umut Kamber, Taige Wang, Amir~S Mohammadi,
  Kenji Watanabe, Takashi Taniguchi, Zlatko Papic, Michael~P Zaletel, and Ali
  Yazdani.
\newblock High-resolution tunneling spectroscopy of fractional quantum hall
  states.
\newblock \emph{arXiv preprint arXiv:2308.05789}, 2023.

\bibitem[Jain and Peterson(2005)]{Jain05}
Jainendra~K. Jain and Michael~R. Peterson.
\newblock Reconstructing the electron in a fractionalized quantum fluid.
\newblock \emph{Phys. Rev. Lett.}, 94:\penalty0 186808, May 2005.
\newblock \doi{10.1103/PhysRevLett.94.186808}.
\newblock URL \url{http://link.aps.org/doi/10.1103/PhysRevLett.94.186808}.

\bibitem[Gattu et~al.(2024)Gattu, Sreejith, and Jain]{Gattu23}
Mytraya Gattu, G.~J. Sreejith, and J.~K. Jain.
\newblock Scanning tunneling microscopy of fractional quantum hall states:
  Spectroscopy of composite-fermion bound states.
\newblock \emph{Phys. Rev. B}, 109:\penalty0 L201123, May 2024.
\newblock \doi{10.1103/PhysRevB.109.L201123}.
\newblock URL \url{https://link.aps.org/doi/10.1103/PhysRevB.109.L201123}.

\bibitem[Pu et~al.(2024{\natexlab{a}})Pu, Balram, Hu, Tsui, He, Regnault,
  Zaletel, Yazdani, and Papić]{Pu24a}
Songyang Pu, Ajit~C. Balram, Yuwen Hu, Yen-Chen Tsui, Minhao He, Nicolas
  Regnault, Michael~P. Zaletel, Ali Yazdani, and Zlatko Papić.
\newblock Fingerprints of composite fermion lambda levels in scanning tunneling
  microscopy.
\newblock \emph{Physical Review B}, 110\penalty0 (8), August
  2024{\natexlab{a}}.
\newblock ISSN 2469-9969.
\newblock \doi{10.1103/physrevb.110.l081107}.
\newblock URL \url{http://dx.doi.org/10.1103/PhysRevB.110.L081107}.

\bibitem[Yue and Stern(2024)]{Yue24}
Xinlei Yue and Ady Stern.
\newblock Electronic excitations in the bulk of fractional quantum hall states.
\newblock \emph{Phys. Rev. B}, 110:\penalty0 115428, Sep 2024.
\newblock \doi{10.1103/PhysRevB.110.115428}.
\newblock URL \url{https://link.aps.org/doi/10.1103/PhysRevB.110.115428}.

\bibitem[Rodr\'{\i}guez et~al.(2012)Rodr\'{\i}guez, Simon, and
  Slingerland]{Rodriguez12}
Iv\'an~D. Rodr\'{\i}guez, Steven~H. Simon, and J.~K. Slingerland.
\newblock Evaluation of ranks of real space and particle entanglement spectra
  for large systems.
\newblock \emph{Phys. Rev. Lett.}, 108:\penalty0 256806, Jun 2012.
\newblock \doi{10.1103/PhysRevLett.108.256806}.
\newblock URL \url{http://link.aps.org/doi/10.1103/PhysRevLett.108.256806}.

\bibitem[Davenport et~al.(2015)Davenport, Rodr\'{\i}guez, Slingerland, and
  Simon]{Davenport15}
Simon~C. Davenport, Iv\'an~D. Rodr\'{\i}guez, J.~K. Slingerland, and Steven~H.
  Simon.
\newblock Composite fermion model for entanglement spectrum of fractional
  quantum {Hall} states.
\newblock \emph{Phys. Rev. B}, 92:\penalty0 115155, Sep 2015.
\newblock \doi{10.1103/PhysRevB.92.115155}.
\newblock URL \url{https://link.aps.org/doi/10.1103/PhysRevB.92.115155}.

\bibitem[Shao et~al.(2015)Shao, Kim, Haldane, and Rezayi]{Shao15}
Junping Shao, Eun-Ah Kim, F.~D.~M. Haldane, and Edward~H. Rezayi.
\newblock Entanglement entropy of the $\ensuremath{\nu}=1/2$ composite fermion
  non-{Fermi} liquid state.
\newblock \emph{Phys. Rev. Lett.}, 114:\penalty0 206402, May 2015.
\newblock \doi{10.1103/PhysRevLett.114.206402}.
\newblock URL \url{http://link.aps.org/doi/10.1103/PhysRevLett.114.206402}.

\bibitem[Newman and Penrose(1966)]{Newman66}
E.~T. Newman and R.~Penrose.
\newblock Note on the bondi-metzner-sachs group.
\newblock \emph{Journal of Mathematical Physics}, 7\penalty0 (5):\penalty0
  863–870, May 1966.
\newblock ISSN 1089-7658.
\newblock \doi{10.1063/1.1931221}.
\newblock URL \url{http://dx.doi.org/10.1063/1.1931221}.

\bibitem[Thorne(1980)]{Thorne80}
Kip~S. Thorne.
\newblock Multipole expansions of gravitational radiation.
\newblock \emph{Reviews of Modern Physics}, 52\penalty0 (2):\penalty0
  299–339, April 1980.
\newblock ISSN 0034-6861.
\newblock \doi{10.1103/revmodphys.52.299}.
\newblock URL \url{http://dx.doi.org/10.1103/RevModPhys.52.299}.

\bibitem[Boyle(2016)]{Boyle16}
Michael Boyle.
\newblock How should spin-weighted spherical functions be defined?
\newblock \emph{Journal of Mathematical Physics}, 57\penalty0 (9), September
  2016.
\newblock ISSN 1089-7658.
\newblock \doi{10.1063/1.4962723}.
\newblock URL \url{http://dx.doi.org/10.1063/1.4962723}.

\bibitem[Wu and Yang(1976)]{Wu76}
Tai~Tsun Wu and Chen~Ning Yang.
\newblock Dirac monopole without strings: Monopole harmonics.
\newblock \emph{Nuclear Physics B}, 107\penalty0 (3):\penalty0 365–380, May
  1976.
\newblock ISSN 0550-3213.
\newblock \doi{10.1016/0550-3213(76)90143-7}.
\newblock URL \url{http://dx.doi.org/10.1016/0550-3213(76)90143-7}.

\bibitem[Wu and Yang(1977)]{Wu77}
Tai~Tsun Wu and Chen~Ning Yang.
\newblock Some properties of monopole harmonics.
\newblock \emph{Physical Review D}, 16\penalty0 (4):\penalty0 1018–1021,
  August 1977.
\newblock ISSN 0556-2821.
\newblock \doi{10.1103/physrevd.16.1018}.
\newblock URL \url{http://dx.doi.org/10.1103/PhysRevD.16.1018}.

\bibitem[Dray(1985)]{Dray85}
Tevian Dray.
\newblock The relationship between monopole harmonics and spin-weighted
  spherical harmonics.
\newblock \emph{Journal of Mathematical Physics}, 26\penalty0 (5):\penalty0
  1030–1033, May 1985.
\newblock ISSN 1089-7658.
\newblock \doi{10.1063/1.526533}.
\newblock URL \url{http://dx.doi.org/10.1063/1.526533}.

\bibitem[Haldane(1983)]{Haldane83}
F.~D.~M. Haldane.
\newblock Fractional quantization of the {Hall} effect: A hierarchy of
  incompressible quantum fluid states.
\newblock \emph{Phys. Rev. Lett.}, 51:\penalty0 605--608, Aug 1983.
\newblock \doi{10.1103/PhysRevLett.51.605}.
\newblock URL \url{http://link.aps.org/doi/10.1103/PhysRevLett.51.605}.

\bibitem[Maciejko et~al.(2013)Maciejko, Hsu, Kivelson, Park, and
  Sondhi]{Maciejko13}
J.~Maciejko, B.~Hsu, S.~A. Kivelson, YeJe Park, and S.~L. Sondhi.
\newblock Field theory of the quantum hall nematic transition.
\newblock \emph{Physical Review B}, 88\penalty0 (12), September 2013.
\newblock ISSN 1550-235X.
\newblock \doi{10.1103/physrevb.88.125137}.
\newblock URL \url{http://dx.doi.org/10.1103/PhysRevB.88.125137}.

\bibitem[Nguyen et~al.(2018)Nguyen, Gromov, and Son]{Nguyen18}
Dung~Xuan Nguyen, Andrey Gromov, and Dam~Thanh Son.
\newblock Fractional quantum hall systems near nematicity: Bimetric theory,
  composite fermions, and dirac brackets.
\newblock \emph{Phys. Rev. B}, 97:\penalty0 195103, May 2018.
\newblock \doi{10.1103/PhysRevB.97.195103}.
\newblock URL \url{https://link.aps.org/doi/10.1103/PhysRevB.97.195103}.

\bibitem[Yang(2020)]{Yang20}
Bo~Yang.
\newblock Microscopic theory for nematic fractional quantum hall effect.
\newblock \emph{Physical Review Research}, 2\penalty0 (3), September 2020.
\newblock ISSN 2643-1564.
\newblock \doi{10.1103/physrevresearch.2.033362}.
\newblock URL \url{http://dx.doi.org/10.1103/PhysRevResearch.2.033362}.

\bibitem[You et~al.(2014)You, Cho, and Fradkin]{You14}
Yizhi You, Gil~Young Cho, and Eduardo Fradkin.
\newblock Theory of nematic fractional quantum hall states.
\newblock \emph{Physical Review X}, 4\penalty0 (4), December 2014.
\newblock ISSN 2160-3308.
\newblock \doi{10.1103/physrevx.4.041050}.
\newblock URL \url{http://dx.doi.org/10.1103/PhysRevX.4.041050}.

\bibitem[Pu et~al.(2024{\natexlab{b}})Pu, Balram, Taylor, Fradkin, and
  Papić]{Pu24}
Songyang Pu, Ajit~C. Balram, Joseph Taylor, Eduardo Fradkin, and Zlatko Papić.
\newblock Microscopic model for fractional quantum hall nematics.
\newblock \emph{Physical Review Letters}, 132\penalty0 (23), June
  2024{\natexlab{b}}.
\newblock ISSN 1079-7114.
\newblock \doi{10.1103/physrevlett.132.236503}.
\newblock URL \url{http://dx.doi.org/10.1103/PhysRevLett.132.236503}.

\bibitem[Chung et~al.(2021)Chung, Villegas~Rosales, Baldwin, Madathil, West,
  Shayegan, and Pfeiffer]{Chung21}
Yoon~Jang Chung, K.~A. Villegas~Rosales, K.~W. Baldwin, P.~T. Madathil, K.~W.
  West, M.~Shayegan, and L.~N. Pfeiffer.
\newblock Ultra-high-quality two-dimensional electron systems.
\newblock \emph{Nature Materials}, Feb 2021.
\newblock ISSN 1476-4660.
\newblock \doi{10.1038/s41563-021-00942-3}.
\newblock URL \url{https://doi.org/10.1038/s41563-021-00942-3}.

\bibitem[Chung et~al.(2022)Chung, Graf, Engel, Rosales, Madathil, Baldwin,
  West, Pfeiffer, and Shayegan]{Chung22}
Yoon~Jang Chung, D.~Graf, L.~W. Engel, K.~A.~Villegas Rosales, P.~T. Madathil,
  K.~W. Baldwin, K.~W. West, L.~N. Pfeiffer, and M.~Shayegan.
\newblock Correlated states of 2{D} electrons near the {Landau} level filling
  $\ensuremath{\nu}=1/7$.
\newblock \emph{Phys. Rev. Lett.}, 128:\penalty0 026802, Jan 2022.
\newblock \doi{10.1103/PhysRevLett.128.026802}.
\newblock URL \url{https://link.aps.org/doi/10.1103/PhysRevLett.128.026802}.

\bibitem[Wigner(1931)]{Wigner31}
Eugen Wigner.
\newblock \emph{Gruppentheorie und ihre Anwendung auf die Quantenmechanik der
  Atomspektren}.
\newblock Vieweg+Teubner Verlag, 1931.
\newblock ISBN 9783663025559.
\newblock \doi{10.1007/978-3-663-02555-9}.
\newblock URL \url{http://dx.doi.org/10.1007/978-3-663-02555-9}.

\bibitem[Sakurai and Napolitano(2020)]{Sakurai20}
J.~J. Sakurai and Jim Napolitano.
\newblock \emph{Modern Quantum Mechanics}.
\newblock Cambridge University Press, September 2020.
\newblock ISBN 9781108473224.
\newblock \doi{10.1017/9781108587280}.
\newblock URL \url{http://dx.doi.org/10.1017/9781108587280}.

\bibitem[SM-()]{SM-Gattu-2024}
Supplemental Materials section includes discussion of how the monopole
  harmonics transform under rotation, derivation of Jain-Kamilla projection in
  the quaternion formulation, and the particle number dependence of the
  magneto-roton dispersion.

\bibitem[Macdonald(1979)]{Macdonald79}
Ian~Grant Macdonald.
\newblock \emph{Symmetric Functions and {Hall} Polynomials}.
\newblock Oxford: Clarendon Press, 1979.

\bibitem[Rehman and Ipsen(2011)]{Rehman11}
Rizwana Rehman and Ilse C.~F. Ipsen.
\newblock Computing characteristic polynomials from eigenvalues.
\newblock \emph{SIAM Journal on Matrix Analysis and Applications}, 32\penalty0
  (1):\penalty0 90–114, January 2011.
\newblock ISSN 1095-7162.
\newblock \doi{10.1137/100788392}.
\newblock URL \url{http://dx.doi.org/10.1137/100788392}.

\bibitem[Feng et~al.(2015)Feng, Wang, Yang, and Jin]{Feng15}
X.~M. Feng, P.~Wang, W.~Yang, and G.~R. Jin.
\newblock High-precision evaluation of wigner's $d$ matrix by exact
  diagonalization.
\newblock \emph{Phys. Rev. E}, 92:\penalty0 043307, Oct 2015.
\newblock \doi{10.1103/PhysRevE.92.043307}.
\newblock URL \url{https://link.aps.org/doi/10.1103/PhysRevE.92.043307}.

\bibitem[Girvin et~al.(1985)Girvin, MacDonald, and Platzman]{Girvin85}
S.~M. Girvin, A.~H. MacDonald, and P.~M. Platzman.
\newblock Collective-excitation gap in the fractional quantum {Hall} effect.
\newblock \emph{Phys. Rev. Lett.}, 54:\penalty0 581--583, Feb 1985.
\newblock \doi{10.1103/PhysRevLett.54.581}.
\newblock URL \url{http://link.aps.org/doi/10.1103/PhysRevLett.54.581}.

\bibitem[Balram et~al.(2024)Balram, Sreejith, and Jain]{Balram24}
Ajit~C. Balram, G.~J. Sreejith, and J.~K. Jain.
\newblock Splitting of the girvin-macdonald-platzman density wave and the
  nature of chiral gravitons in the fractional quantum hall effect.
\newblock \emph{Phys. Rev. Lett.}, 133:\penalty0 246605, Dec 2024.
\newblock \doi{10.1103/PhysRevLett.133.246605}.
\newblock URL \url{https://link.aps.org/doi/10.1103/PhysRevLett.133.246605}.

\bibitem[Simon and Halperin(1993)]{Simon93}
Steven~H. Simon and Bertrand~I. Halperin.
\newblock Finite-wave-vector electromagnetic response of fractional quantized
  {Hall} states.
\newblock \emph{Phys. Rev. B}, 48:\penalty0 17368--17387, Dec 1993.
\newblock \doi{10.1103/PhysRevB.48.17368}.
\newblock URL \url{http://link.aps.org/doi/10.1103/PhysRevB.48.17368}.

\bibitem[Golkar et~al.(2016)Golkar, Nguyen, Roberts, and Son]{Golkar16}
Siavash Golkar, Dung~Xuan Nguyen, Matthew~M. Roberts, and Dam~Thanh Son.
\newblock Higher-spin theory of the magnetorotons.
\newblock \emph{Phys. Rev. Lett.}, 117:\penalty0 216403, Nov 2016.
\newblock \doi{10.1103/PhysRevLett.117.216403}.
\newblock URL \url{http://link.aps.org/doi/10.1103/PhysRevLett.117.216403}.

\bibitem[Park and Jain(2000)]{Park00}
K.~Park and J.~K. Jain.
\newblock Two-roton bound state in the fractional quantum {Hall} effect.
\newblock \emph{Phys. Rev. Lett.}, 84:\penalty0 5576--5579, Jun 2000.
\newblock \doi{10.1103/PhysRevLett.84.5576}.
\newblock URL \url{http://link.aps.org/doi/10.1103/PhysRevLett.84.5576}.

\bibitem[Balram and Pu(2017)]{Balram17c}
Ajit~C Balram and Songyang Pu.
\newblock Positions of the magnetoroton minima in the fractional quantum hall
  effect.
\newblock \emph{The European Physical Journal B: Condensed Matter and Complex
  Systems}, 90\penalty0 (6):\penalty0 124, 2017.

\bibitem[Read and Rezayi(1999)]{Read99}
N.~Read and E.~Rezayi.
\newblock Beyond paired quantum {Hall} states: Parafermions and incompressible
  states in the first excited {Landau} level.
\newblock \emph{Phys. Rev. B}, 59:\penalty0 8084--8092, Mar 1999.
\newblock \doi{10.1103/PhysRevB.59.8084}.
\newblock URL \url{http://link.aps.org/doi/10.1103/PhysRevB.59.8084}.

\bibitem[Bonderson et~al.(2012)Bonderson, Feiguin, M\"oller, and
  Slingerland]{Bonderson12}
Parsa Bonderson, Adrian~E. Feiguin, Gunnar M\"oller, and J.~K. Slingerland.
\newblock Competing topological orders in the $\ensuremath{\nu}=12/5$ quantum
  {Hall} state.
\newblock \emph{Phys. Rev. Lett.}, 108:\penalty0 036806, Jan 2012.
\newblock \doi{10.1103/PhysRevLett.108.036806}.
\newblock URL \url{http://link.aps.org/doi/10.1103/PhysRevLett.108.036806}.

\bibitem[Sreejith et~al.(2013)Sreejith, Wu, W\'ojs, and Jain]{Sreejith13}
G.~J. Sreejith, Ying-Hai Wu, A.~W\'ojs, and J.~K. Jain.
\newblock Tripartite composite fermion states.
\newblock \emph{Phys. Rev. B}, 87:\penalty0 245125, Jun 2013.
\newblock \doi{10.1103/PhysRevB.87.245125}.
\newblock URL \url{http://link.aps.org/doi/10.1103/PhysRevB.87.245125}.

\bibitem[Repellin et~al.(2015)Repellin, Neupert, Bernevig, and
  Regnault]{Repellin15}
C\'ecile Repellin, Titus Neupert, B.~Andrei Bernevig, and Nicolas Regnault.
\newblock Projective construction of the ${\mathbb{z}}_{k}$ {Read}-{Rezayi}
  fractional quantum {Hall} states and their excitations on the torus geometry.
\newblock \emph{Phys. Rev. B}, 92:\penalty0 115128, Sep 2015.
\newblock \doi{10.1103/PhysRevB.92.115128}.
\newblock URL \url{https://link.aps.org/doi/10.1103/PhysRevB.92.115128}.

\bibitem[Ku\ifmmode~\acute{s}\else \'{s}\fi{}mierz and
  W\'ojs(2018)]{Kusmierz18}
Bartosz Ku\ifmmode~\acute{s}\else \'{s}\fi{}mierz and Arkadiusz W\'ojs.
\newblock Emergence of jack ground states from two-body pseudopotentials in
  fractional quantum {Hall} systems.
\newblock \emph{Phys. Rev. B}, 97:\penalty0 245125, Jun 2018.
\newblock \doi{10.1103/PhysRevB.97.245125}.
\newblock URL \url{https://link.aps.org/doi/10.1103/PhysRevB.97.245125}.

\bibitem[Balram(2022)]{Balram22a}
Ajit~C Balram.
\newblock Transitions from abelian composite fermion to non-abelian parton
  fractional quantum hall states in the zeroth landau level of bilayer
  graphene.
\newblock \emph{Physical Review B}, 105\penalty0 (12):\penalty0 L121406, 2022.

\bibitem[Danisch and Krumbiegel(2021)]{Danisch21}
Simon Danisch and Julius Krumbiegel.
\newblock Makie.jl: Flexible high-performance data visualization for julia.
\newblock \emph{Journal of Open Source Software}, 6\penalty0 (65):\penalty0
  3349, September 2021.
\newblock ISSN 2475-9066.
\newblock \doi{10.21105/joss.03349}.
\newblock URL \url{http://dx.doi.org/10.21105/joss.03349}.

\end{thebibliography}


\begin{thebibliography}{17}
\providecommand{\natexlab}[1]{#1}
\providecommand{\url}[1]{\texttt{#1}}
\expandafter\ifx\csname urlstyle\endcsname\relax
  \providecommand{\doi}[1]{doi: #1}\else
  \providecommand{\doi}{doi: \begingroup \urlstyle{rm}\Url}\fi

\bibitem[Wu and Yang(1976)]{Wu76}
Tai~Tsun Wu and Chen~Ning Yang.
\newblock Dirac monopole without strings: Monopole harmonics.
\newblock \emph{Nuclear Physics B}, 107\penalty0 (3):\penalty0 365–380, May
  1976.
\newblock ISSN 0550-3213.
\newblock \doi{10.1016/0550-3213(76)90143-7}.
\newblock URL \url{http://dx.doi.org/10.1016/0550-3213(76)90143-7}.

\bibitem[Wu and Yang(1977)]{Wu77}
Tai~Tsun Wu and Chen~Ning Yang.
\newblock Some properties of monopole harmonics.
\newblock \emph{Physical Review D}, 16\penalty0 (4):\penalty0 1018–1021,
  August 1977.
\newblock ISSN 0556-2821.
\newblock \doi{10.1103/physrevd.16.1018}.
\newblock URL \url{http://dx.doi.org/10.1103/PhysRevD.16.1018}.

\bibitem[Jain(2007)]{Jain07}
J.~K. Jain.
\newblock \emph{Composite Fermions}.
\newblock Cambridge University Press, New York, US, 2007.

\bibitem[Wigner(1931)]{Wigner31}
Eugen Wigner.
\newblock \emph{Gruppentheorie und ihre Anwendung auf die Quantenmechanik der
  Atomspektren}.
\newblock Vieweg+Teubner Verlag, 1931.
\newblock ISBN 9783663025559.
\newblock \doi{10.1007/978-3-663-02555-9}.
\newblock URL \url{http://dx.doi.org/10.1007/978-3-663-02555-9}.

\bibitem[Sakurai and Napolitano(2020)]{Sakurai20}
J.~J. Sakurai and Jim Napolitano.
\newblock \emph{Modern Quantum Mechanics}.
\newblock Cambridge University Press, September 2020.
\newblock ISBN 9781108473224.
\newblock \doi{10.1017/9781108587280}.
\newblock URL \url{http://dx.doi.org/10.1017/9781108587280}.

\bibitem[Zhang et~al.(2016)Zhang, Wójs, and Jain]{Zhang16}
Yuhe Zhang, A.~Wójs, and J.~K. Jain.
\newblock Landau-level mixing and particle-hole symmetry breaking for spin
  transitions in the fractional quantum hall effect.
\newblock \emph{Physical Review Letters}, 117\penalty0 (11), September 2016.
\newblock ISSN 1079-7114.
\newblock \doi{10.1103/physrevlett.117.116803}.
\newblock URL \url{http://dx.doi.org/10.1103/PhysRevLett.117.116803}.

\bibitem[Zhao et~al.(2018)Zhao, Zhang, and Jain]{Zhao18}
Jianyun Zhao, Yuhe Zhang, and J.~K. Jain.
\newblock Crystallization in the fractional quantum hall regime induced by
  landau-level mixing.
\newblock \emph{Physical Review Letters}, 121\penalty0 (11), September 2018.
\newblock ISSN 1079-7114.
\newblock \doi{10.1103/physrevlett.121.116802}.
\newblock URL \url{http://dx.doi.org/10.1103/PhysRevLett.121.116802}.

\bibitem[Jain and Kamilla(1997)]{Jain97}
J.~K. Jain and R.~K. Kamilla.
\newblock Composite fermions in the {Hilbert} space of the lowest electronic
  {Landau} level.
\newblock \emph{Int. J. Mod. Phys. B}, 11\penalty0 (22):\penalty0 2621--2660,
  1997.
\newblock \doi{10.1142/S0217979297001301}.

\bibitem[Davenport and Simon(2012)]{Davenport12}
Simon~C. Davenport and Steven~H. Simon.
\newblock Spinful composite fermions in a negative effective field.
\newblock \emph{Phys. Rev. B}, 85:\penalty0 245303, Jun 2012.
\newblock \doi{10.1103/PhysRevB.85.245303}.
\newblock URL \url{http://link.aps.org/doi/10.1103/PhysRevB.85.245303}.

\bibitem[Dou()]{DoubleFloats}
{DoubleFloats}.jl: math with more good bits.
\newblock URL \url{https://github.com/JuliaMath/DoubleFloats.jl}.

\bibitem[Macdonald(1979)]{Macdonald79}
Ian~Grant Macdonald.
\newblock \emph{Symmetric Functions and {Hall} Polynomials}.
\newblock Oxford: Clarendon Press, 1979.

\bibitem[Feng et~al.(2015)Feng, Wang, Yang, and Jin]{Feng15}
X.~M. Feng, P.~Wang, W.~Yang, and G.~R. Jin.
\newblock High-precision evaluation of wigner's $d$ matrix by exact
  diagonalization.
\newblock \emph{Phys. Rev. E}, 92:\penalty0 043307, Oct 2015.
\newblock \doi{10.1103/PhysRevE.92.043307}.
\newblock URL \url{https://link.aps.org/doi/10.1103/PhysRevE.92.043307}.

\bibitem[Mandal and Jain(2002)]{Mandal02}
Sudhansu~S. Mandal and Jainendra~K. Jain.
\newblock Theoretical search for the nested quantum {Hall} effect of composite
  fermions.
\newblock \emph{Phys. Rev. B}, 66:\penalty0 155302, Oct 2002.
\newblock \doi{10.1103/PhysRevB.66.155302}.
\newblock URL \url{http://link.aps.org/doi/10.1103/PhysRevB.66.155302}.

\bibitem[Harville(1997)]{Harville97}
David~A. Harville.
\newblock \emph{Matrix Algebra From a Statistician’s Perspective}.
\newblock Springer New York, 1997.
\newblock ISBN 9780387226774.
\newblock \doi{10.1007/b98818}.
\newblock URL \url{http://dx.doi.org/10.1007/b98818}.

\bibitem[Balram et~al.(2013)Balram, W\'ojs, and Jain]{Balram13}
Ajit~C. Balram, Arkadiusz W\'ojs, and Jainendra~K. Jain.
\newblock State counting for excited bands of the fractional quantum {Hall}
  effect: Exclusion rules for bound excitons.
\newblock \emph{Phys. Rev. B}, 88:\penalty0 205312, Nov 2013.
\newblock \doi{10.1103/PhysRevB.88.205312}.
\newblock URL \url{http://link.aps.org/doi/10.1103/PhysRevB.88.205312}.

\bibitem[Ramachandran and Tsokos(2021)]{Ramachandran21}
Kandethody~M. Ramachandran and Chris~P. Tsokos.
\newblock \emph{Bayesian estimation and inference}, page 415–459.
\newblock Elsevier, 2021.
\newblock ISBN 9780128178157.
\newblock \doi{10.1016/b978-0-12-817815-7.00010-5}.
\newblock URL \url{http://dx.doi.org/10.1016/B978-0-12-817815-7.00010-5}.

\bibitem[Morf et~al.(1986)Morf, d'Ambrumenil, and Halperin]{Morf86}
R.~Morf, N.~d'Ambrumenil, and B.~I. Halperin.
\newblock Microscopic wave functions for the fractional quantized {Hall} states
  at $\ensuremath{\nu}=\frac{2}{5} \mathrm{and} \frac{2}{7}$.
\newblock \emph{Phys. Rev. B}, 34:\penalty0 3037--3040, Aug 1986.
\newblock \doi{10.1103/PhysRevB.34.3037}.
\newblock URL \url{http://link.aps.org/doi/10.1103/PhysRevB.34.3037}.

\end{thebibliography}
\end{document}

\title{Supplementary Material for ``Unlocking new regimes in fractional quantum Hall effect with quaternions"}
\author{Mytraya Gattu and J. K. Jain}
\affiliation{Department of Physics, 104 Davey Lab, Pennsylvania State University, University Park, Pennsylvania 16802, USA}
\date{\today}
\maketitle
In Sec.~\ref{appendix-monopole-harmonics-under-rotation}, we provide a detailed analysis of the transformation properties of monopole harmonics under rotation, when expressed in terms of the spinor variables. Specifically, we demonstrate that the phase factor they acquire under rotations is discontinuous at points mapping to the poles. In Sec.~\ref{appendix-jk-projection-quaternions}, we derive the expression for Jain-Kamilla projection in the quaternion representation and discuss how it leads to a much more efficient evaluation of the Jain-Kamilla wave functions. Finally, in Sec.~\ref{appendix-roton-dispersion}, we provide dispersions for the magneto-roton modes at several Jain fractions.

\section{Monopole harmonics under rotation}\label{appendix-monopole-harmonics-under-rotation}

Eq.~2 of the main text (Eq.~\ref{eq:appendix-monopole-harmonics-under-rotation} below), which shows how the monopole harmonics $Y_{Q, l, m}(\theta, \phi)$ transform under rotations described by the Euler angles $\alpha, \beta, \gamma$ about the $z-y-z$ axes, was first derived in Ref.~\cite{Wu76,Wu77}. We include an alternate derivation below for completeness and for the reader's convenience. Through examples, we show that the phase $Q\psi'$ in Eq.~\ref{eq:appendix-monopole-harmonics-under-rotation} acquired by the monopole harmonics under a rotation is in general discontinuous at the points that go to the poles under the rotation.

In the spherical geometry, the single particle eigenstates in a uniform radial magnetic field produced by a monopole of strength $Q$ are the monopole harmonics $Y_{Q, l, m}(\Omega)$. Here, $l = \abs{Q}, \abs{Q}+1, \dots$ is the angular momentum, $m = -l, -l+1, \dots, l$ is the azimuthal quantum number, and $\Omega = (\theta, \phi)$ is the position in spherical coordinates. The monopole harmonics $Y_{Q, l, m}(\theta, \phi)$ are defined as polynomials in the spinor variables $u = \cos (\theta/2) e^{\imath \phi / 2}$ and $v = \sin (\theta / 2) e^{-\imath \phi / 2}$ as~\cite{Wu76, Jain07}
\begin{equation} \label{eq:appendix-monopole-harmonics-sphere}
\begin{split}
 &Y_{Q,l,m}(\theta, \phi) = N_{Q,l,m}(-1)^{l-m}v^{Q-m}u^{Q+m}\\
 &\times \sum_{s}(-1)^{s}\binom{l-Q}{s}\binom{l+Q}{l-m-s}(\conj{v}v)^{l-Q-s}(\conj{u}u)^{s}\\
  &N_{Q, l, m} = \sqrt{\frac{2l+1}{4\pi}\frac{(l+m)!(l-m)!}{(l+Q)!(l-Q)!}}
 \end{split}
\end{equation}

From the rotational symmetry of the spherical geometry and from general symmetry considerations~\cite{Wigner31}, it follows that we should be able to write
\begin{equation}\label{eq:general-transformation-law}
    \begin{split}
        &Y_{Q, l, m}(\theta^{\prime}, \phi^{\prime})= \\
        &e^{\imath \Gamma}\sum_{m^{\prime}=-l}^{l}U^{l}_{m, m^{\prime}}(\mathcal{R}(\alpha, \beta, \gamma))Y_{Q, l, m^{\prime}}(\theta, \phi)
    \end{split}
\end{equation}
at every point $(\theta^{\prime}, \phi^{\prime})$ which is related to the point $(\theta, \phi)$ by the rotation $\mathcal{R}(\alpha, \beta, \gamma)$. Here, $U^{l}_{m, m^{\prime}}(\mathcal{R}(\alpha, \beta, \gamma))$ is a unitary matrix generating the rotation and $\Gamma$ is a phase which, in general, depends on both $(\theta, \phi)$ and $\mathcal{R}(\alpha, \beta, \gamma)$. Now, the question is: Can we find an explicit form for $U^{l}_{m, m^{\prime}}(\mathcal{R}(\alpha, \beta, \gamma))$ and $\Gamma$ to be able to exploit the rotational invariance?

We can in principle, relate $Y_{Q, l, m}(\theta^{\prime}, \phi^{\prime})$ to $Y_{Q, l, m}(\theta, \phi)$ by expressing the spinor variables $u^{\prime} = \cos(\theta^{\prime}/2)e^{\imath \phi^{\prime}/2}$ and $v^{\prime}=\sin(\theta^{\prime}/2)e^{-\imath \phi^{\prime}/2}$ appearing in Eq.~\ref{eq:appendix-monopole-harmonics-sphere} in terms of the original spinor variables $u = \cos(\theta/2)e^{\imath \phi/2}$ and $v^{\prime}=\sin(\theta/2)e^{-\imath \phi/2}$. To do so would require solving 
\begin{equation}\label{eq:sphere-rotation}
    \begin{split}
        &\left(\begin{matrix}
            \sin \theta^{\prime} \cos \phi^{\prime}\\
            \sin \theta^{\prime} \sin \phi^{\prime}\\
            \cos \theta^{\prime}\\
        \end{matrix}\right) = \left(\begin{matrix}
            \cos \alpha & -\sin \alpha & 0 \\
            \sin \alpha & \cos \alpha & 0 \\
            0 & 0 & 1 \\
        \end{matrix}\right)\times \\
        &\left(\begin{matrix}
            \cos \beta & 0 & \sin \beta \\
            0 & 1 & 0 \\
            -\sin \beta & 0 & \cos \beta \\
        \end{matrix}\right)\left(\begin{matrix}
            \cos \gamma & -\sin \gamma & 0 \\
            \sin \gamma & \cos \gamma & 0 \\
            0 & 0 & 1 \\
        \end{matrix}\right)\left(\begin{matrix}
            \sin \theta \cos \phi\\
            \sin \theta \sin \phi\\
            \cos \theta\\
        \end{matrix}\right)
    \end{split}.
\end{equation}
for $\cos(\theta^{\prime}/2), \sin(\theta^{\prime}/2), \cos(\phi^{\prime}/2)$ and $\sin(\phi^{\prime}/2)$. Except for the most straightforward rotations (i.e., the ones about the $z$-axis alone), solving for the new spinor variables would require a careful consideration of branch cuts. 

A more practical and insightful approach can be taken by noting that, mathematically, the monopole harmonics $Y_{Q, l, m}(\theta, \phi)$ can be related to a row of the unitary Wigner-$D$ matrices that generate rotations in three dimensions as~\cite{Wu76}:
\begin{equation}\label{eq:monopole-harmonics-wigner-D-matrices-relation}
\begin{split}
&Y_{Q, l, m}(\theta, \phi) = \sqrt{\frac{2l+1}{4\pi}}D^{l}_{Q, m}(0,\theta,-\phi)\\
\end{split}
\end{equation}
Here, $D^{l}_{m, m^{\prime}}(\alpha, \beta, \gamma)$ is the Wigner-$D$ matrix generating an active rotation described by the Euler angles $\alpha,\beta,\gamma$ about the $z-y-z$ axes of a spin-$l$ particle~\cite{Sakurai20}:
\begin{equation}\label{eq:wigner-D-matrix}
 D^{l}_{m, m^{\prime}}(\alpha, \beta, \gamma) = \bra{l, m}e^{-\imath \alpha J_{z}}e^{-\imath \beta J_{y}}e^{-\imath \gamma J_{z}}\ket{l, m^{\prime}}    
\end{equation}
Here, $J_{x}, J_{y}, J_{z}$ are the angular momentum operators and $\overlap{\theta, \phi}{l, m} = Y_{l, m}(\theta, \phi)$ are the spherical harmonics of angular momentum $l$. 
[We will follow the above convention for the Euler angles and refer to such rotations as $\mathcal{R}(\alpha, \beta, \gamma)$, with $\mathcal{R}^{-1}(\alpha, \beta, \gamma)=\mathcal{R}(-\gamma, -\beta, -\alpha)$.]
Therefore, to relate the monopole harmonics $Y_{Q, l, m}(\theta^{\prime}, \phi^{\prime})$ to the monopole harmonics $Y_{Q, l, m}(\theta, \phi)$, all we need to do is to express the rotation $\mathcal{R}(0, \theta^{\prime}, -\phi^{\prime})$ in terms of the rotation $\mathcal{R}(0, \theta, -\phi)$. If we can do this, we can use the transformation rule for the Wigner-$D$ matrices under the composition of rotations, $D^{l}_{m_{1}, m_{2}}(\mathcal{R}_{1}\cdot \mathcal{R}_{2}) = \sum_{m^{\prime}=-l}^{l}D^{l}_{m_{1}, m^{\prime}}(\mathcal{R}_{1})D^{l}_{m^{\prime}, m_{2}}(\mathcal{R}_{2})$ to construct the transformation rule for the monopole harmonics under rotations.

To proceed further, consider the following thought exercise. First, what does the inverse of the rotation $\mathcal{R}(0, \theta, -\phi)$ do? Its inverse $\mathcal{R}(\phi, -\theta, 0)$, takes the north pole to the point $(\theta, \pi + \phi)$. If we now apply another rotation $\mathcal{R}(0, 0, \pi)$, we can bring the north pole under the combined action to $(\theta, \phi)$. Now, we know that, under $\mathcal{R}(\alpha, \beta, \gamma)$, $(\theta, \phi)$ goes to $(\theta^{\prime}, \phi^{\prime})$ and therefore, under the combined rotation $\mathcal{R}(\alpha, \beta, \gamma)\cdot \mathcal{R}(0, 0, \pi)\cdot \mathcal{R}(\phi, -\theta, 0)$, the north pole goes to $(\theta^{\prime}, \phi^{\prime})$. But, we also know that we can bring the north pole to $(\theta^{\prime}, \phi^{\prime})$, by the rotation $\mathcal{R}(0, 0, \pi)\cdot \mathcal{R}(\phi^{\prime}, -\theta^{\prime}, 0)$. Accounting for the fact that the north pole remains invariant under arbitrary rotations about the $z$-axis (which we take to be equal to $\mathcal{R}(0, 0, \psi^{\prime})$), we must have:
\begin{align*}
    &\mathcal{R}(0, 0, \pi)\cdot \mathcal{R}(\phi^{\prime}, -\theta^{\prime}, 0) \cdot \mathcal{R}(0, 0, \psi^{\prime})\\
    &= \mathcal{R}(\alpha, \beta, \gamma)\cdot \mathcal{R}(0, 0, \pi)\cdot \mathcal{R}(\phi, -\theta, 0)
\end{align*}
Or, equivalently we can invert both sides to get:
\begin{align*}
    &\Leftrightarrow \mathcal{R}(-\psi^{\prime}, 0, 0)\cdot \mathcal{R}(0, \theta^{\prime}, -\phi^{\prime}) \cdot \mathcal{R}(-\pi,0,0)\\
    &= \mathcal{R}(0,\theta,-\phi)\cdot \mathcal{R}(-\pi,0,0)\cdot \mathcal{R}(-\gamma, -\beta, -\alpha)
\end{align*}
We have thus managed to relate the rotation $\mathcal{R}(0, \theta, -\phi)$ to the rotation $\mathcal{R}(0, \theta^{\prime}, -\phi^{\prime})$. We can use this relation to find the transformation rule for the monopole harmonics. From Eq.~\ref{eq:monopole-harmonics-wigner-D-matrices-relation} and Eq.~\ref{eq:wigner-D-matrix}, the L.H.S. gives us:
\begin{align*}
    &D^{l}_{Q, m}(\mathcal{R}(-\psi^{\prime}, 0, 0)\cdot \mathcal{R}(0, \theta^{\prime}, -\phi^{\prime}) \cdot \mathcal{R}(-\pi,0,0))\\
    &=\bra{l,Q}e^{\imath \psi^{\prime} J_{z}}\mathcal{R}(0,\theta^{\prime},-\phi^{\prime})e^{\imath \pi J_{z}}\ket{l,m}\\
    &=\sqrt{\frac{4\pi}{2l+1}}(-1)^{m}e^{\imath Q \psi^{\prime}}Y_{Q,l,m}(\theta^{\prime}, \phi^{\prime}).
\end{align*}
Similarly, the R.H.S. gives us (recall, $D^{l}_{m, m^{\prime}}(\mathcal{R}) = \left[D^{l}_{m^{\prime}, m}(\mathcal{R}^{-1})\right]^{\star}$):
\begin{align*}
    &D^{l}_{Q, m}(\mathcal{R}(0,\theta,-\phi)\cdot \mathcal{R}(-\pi,0,0)\cdot \mathcal{R}(-\gamma, -\beta, -\alpha)) \\
    &=\sum_{m^{\prime}}(-1)^{m^{\prime}}D^{l}_{Q, m^{\prime}}(\mathcal{R}(0,\theta,-\phi))D^{l}_{m^{\prime},m}(\mathcal{R}(-\gamma, -\beta, -\alpha)) \\
    &=\sqrt{\frac{4\pi}{2l+1}}\sum_{m^{\prime}}(-1)^{m^{\prime}}Y_{Q,l,m^{\prime}}(\theta, \phi)[D^{l}_{m, m^{\prime}}(\mathcal{R}(\alpha, \beta, \gamma))]^{\star}.
\end{align*}

Therefore, we find that under a rotation $\mathcal{R}(\alpha, \beta, \gamma)$, which takes the point $(\theta, \phi)$ to the point $(\theta^{\prime}, \phi^{\prime})$, the monopole harmonics $Y_{Q, l, m}(\theta, \phi)$ transform as:
\begin{equation}\label{eq:appendix-monopole-harmonics-under-rotation}
    \begin{split}
   &Y_{Q, l, m}(\theta^{\prime}, \phi^{\prime})\\
        &= e^{-\imath Q \psi^{\prime}}\sum_{m^{\prime}=-l}^{l}(-1)^{m-m^{\prime}}[D^{l}_{m, m^{\prime}}(\alpha, \beta, \gamma)]^{\star}Y_{Q, l, m^{\prime}}(\theta, \phi).
    \end{split}
\end{equation}
We thus find that in Eq.~\ref{eq:general-transformation-law}, the unitary matrix $U^{l}_{m, m^{\prime}}$ is equal to $(-1)^{m-m^{\prime}}[D^{l}_{m, m^{\prime}}(\alpha, \beta, \gamma)]^{\star}$ and the phase $\Gamma$ is equal to $-Q \psi^{\prime}$ with $\psi^{\prime}$ determined numerically from:
\begin{equation}
    \mathcal{R}(0, 0, \psi^{\prime}) = \mathcal{R}(0, \theta^{\prime}, -\phi^{\prime})\cdot \mathcal{R}(\alpha-\pi, \beta, \gamma + \pi) \cdot \mathcal{R}(\phi, -\theta, 0)
\end{equation}

\begin{figure}
    \includegraphics[width=\columnwidth]{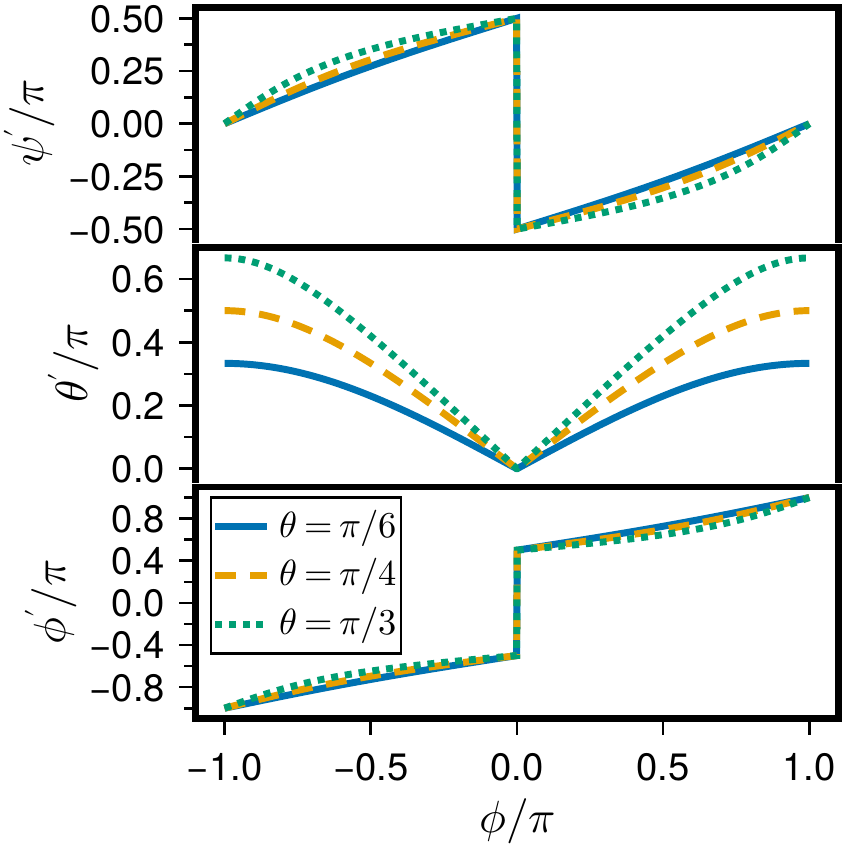}
    \caption{In the top row, for rotations $\mathcal{R}(\alpha = 0, \beta = -\theta, \gamma = 0)$ that take the points $\Omega = (\theta, 0)$ to the north pole, we have plotted the angle $\psi^{\prime}$ as a function of the original azimuthal angle $\phi$ for points initially along the $\theta$ latitude. (The phase acquired by monopole harmonics $Y_{Q, l, m}(\theta, \phi)$ under rotation is $Q\psi^{\prime}$.) In the middle and bottom rows, we plot, for reference, $\theta^{\prime}$ and $\phi^{\prime}$ coordinates of these points after rotation. 
    }
    \label{fig:phase-under-rotation-special} 
\end{figure}

\begin{figure}
    \includegraphics[width=0.95\columnwidth]{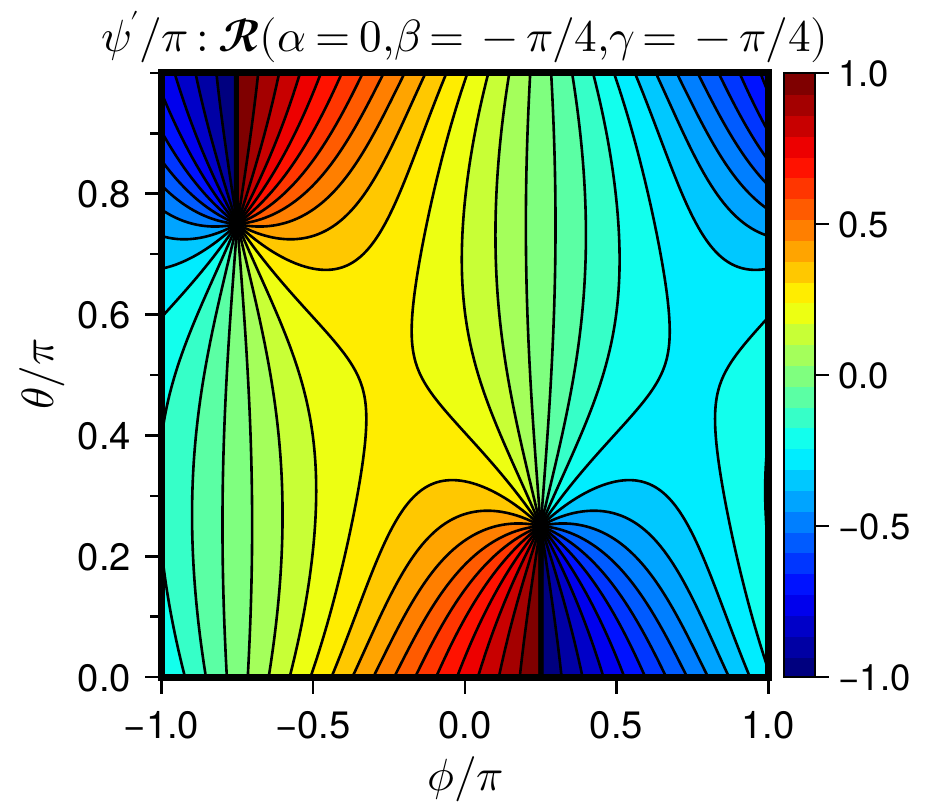}
    \includegraphics[width=0.95\columnwidth]{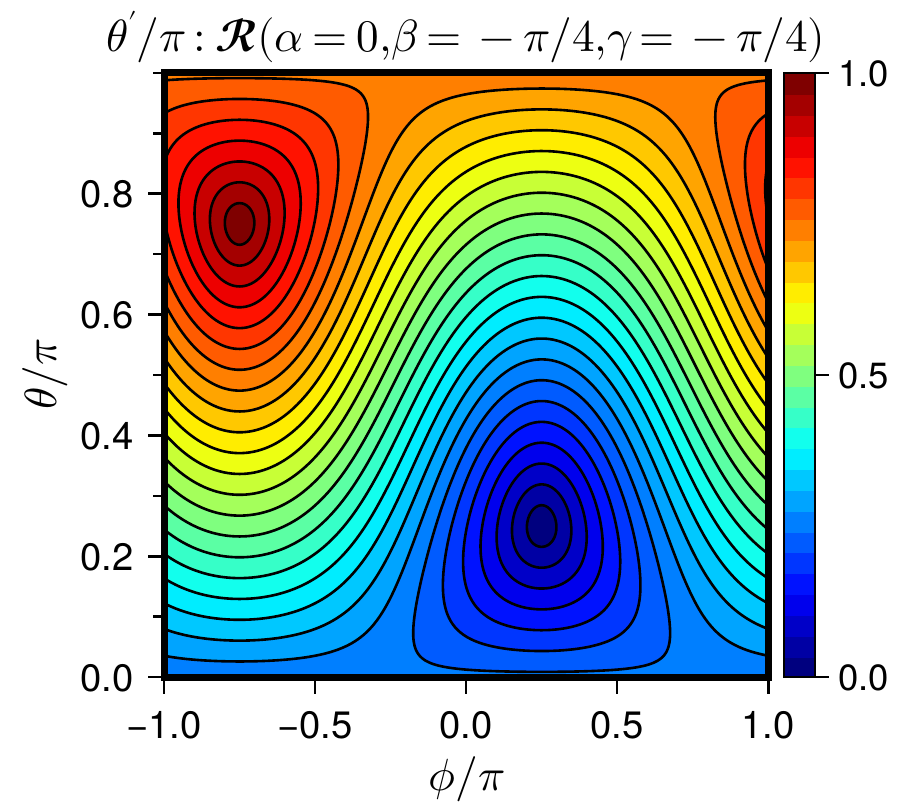}
    \includegraphics[width=0.95\columnwidth]{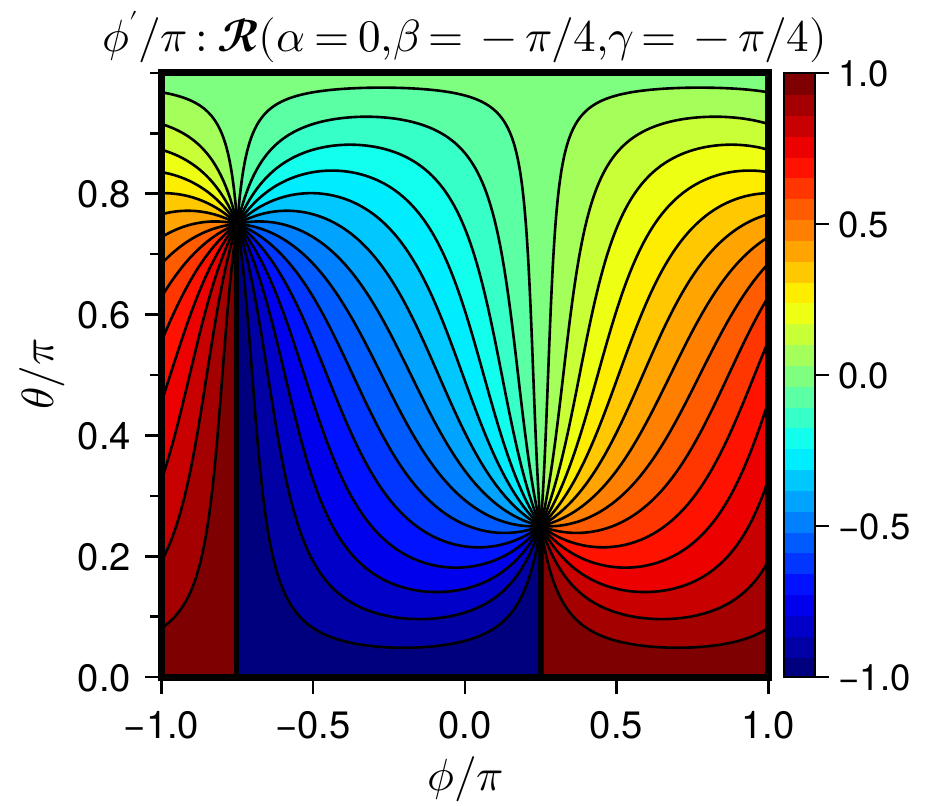}
    \caption{For the rotation $\mathcal{R}(\alpha = 0, \beta = -\pi/4, \gamma = -\pi/4)$, we plot $\psi^{\prime}$, $\theta^{\prime}$ and $\phi^{\prime}$ as a function of the initial coordinates $(\theta, \phi)$ over the entire sphere.
    }
    \label{fig:phase-under-rotation-generic} 
\end{figure}

We present solutions for $\psi^{\prime}$ in Fig.~\ref{fig:phase-under-rotation-special} and Fig.~\ref{fig:phase-under-rotation-generic} for some parameter choices, which demonstrate that $\psi^{\prime}$ in general depends on $(\theta, \phi)$ and $\mathcal{R}(\alpha, \beta, \gamma)$ in a complicated manner and is discontinuous at points which go to the poles under the rotation $\mathcal{R}(\alpha, \beta, \gamma)$. A priori, we expect such a discontinuity at points that go to the poles due to the presence of Dirac strings (recall that the Dirac strings are fixed to the poles of the co-ordinate system).

\section{Jain-Kamilla projection in terms of unit quaternions}\label{appendix-jk-projection-quaternions}

In this section, we first review the traditional formulation of the Jain-Kamilla (JK) projection method, which provides a variational ansatz for the lowest Landau level (LLL) projected composite fermion (CF) wave functions. We then derive a novel reformulation of the JK projection employing the quaternion formulation of the monopole harmonics. We discuss its practical implementation and show that the quaternion formulation dramatically increases the computational speed and accuracy of JK projected wave functions, making it possible to study Jain FQH states at fillings $\nu = 1/3, 2/5, \dots, 20/41$ and $\nu = 2/3, 3/5, \cdots, 19/37, 20/39$.

A crucial ingredient of the wave functions of CFs is the $\LLL$ operator in their definition (for a brief review, see Sec.~\ref{appendix-roton-dispersion} or Ref.~\cite{Jain07}). Therefore, performing the LLL projection efficiently is vital in the numerical studies of CFs. (The unprojected wave functions have occupation of higher Landau levels but they \addMG{do} not describe the physics of Landau level mixing. Even for describing the physics of LL mixing, one begins with the LLL wave function~\cite{Zhang16,Zhao18}.) To this end, Jain and Kamilla~\cite{Jain97} proposed an ansatz for the LLL projected CF wave functions:
\begin{equation}
    \begin{split}
        &\LLL \det{\left[Y_{Q^{\star}, l_{j}, m_{j}}(\Omega_{i})\right]}\prod_{a<b=1}^{N}(u_{a}v_{b}-u_{b}v_{a})^{2}\\
        &\rightarrow \det \left[\LLL Y_{Q^{\star}, l_{j}, m_{j}}(\Omega_{i})\prod_{k \neq i}(u_{i}v_{k}-u_{k}v_{i})\right]
    \end{split}
\end{equation}
In the JK projected wave functions, the problem of projecting an $N$-particle wave function into the LLL, which has an $\mathcal{O}(N^{N})$ complexity (recall that a Landau level has a macroscopic degeneracy), is reduced to projecting each element $Y_{Q^{\star}, l_{j}, m_{j}}(\Omega_{i})\prod_{k \neq i}(u_{i}v_{k}-u_{k}v_{i})$ of a Slater determinant that contains at most one particle outside the LLL, which has an overall complexity $\mathcal{O}(N^{2-3})$. This makes it possible to study systems with $N$ up to a few hundred electrons, thus allowing for a quantitative understanding of the thermodynamic properties of many FQH states.

\begin{figure}
    \includegraphics[width=0.68\columnwidth]{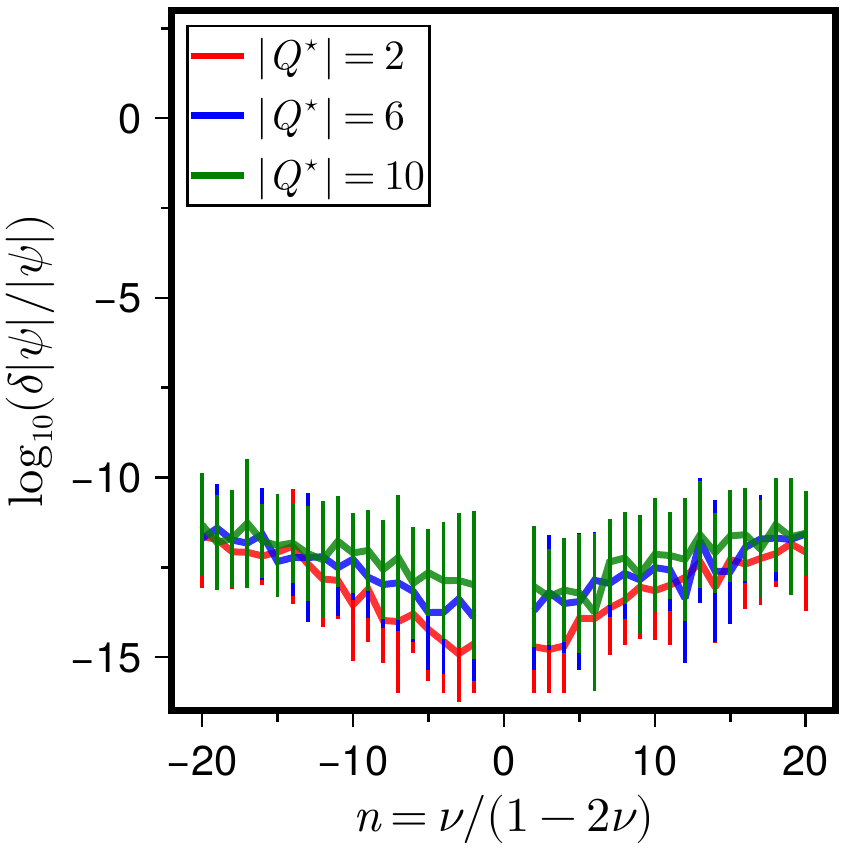}
    \includegraphics[width=0.68\columnwidth]{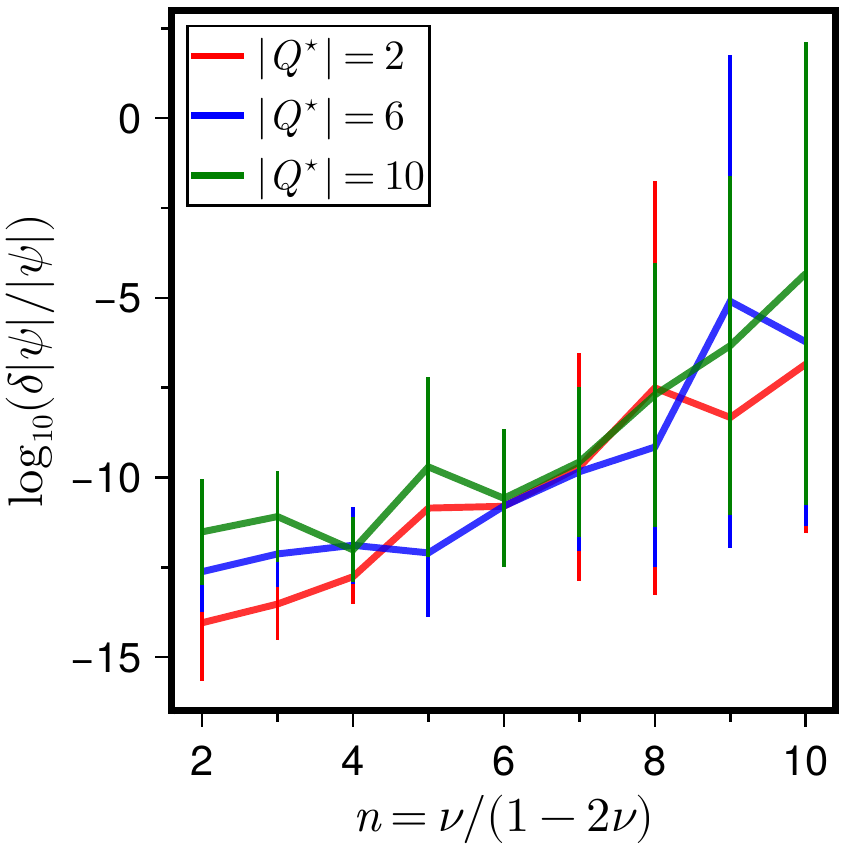}
    \includegraphics[width=0.68\columnwidth]{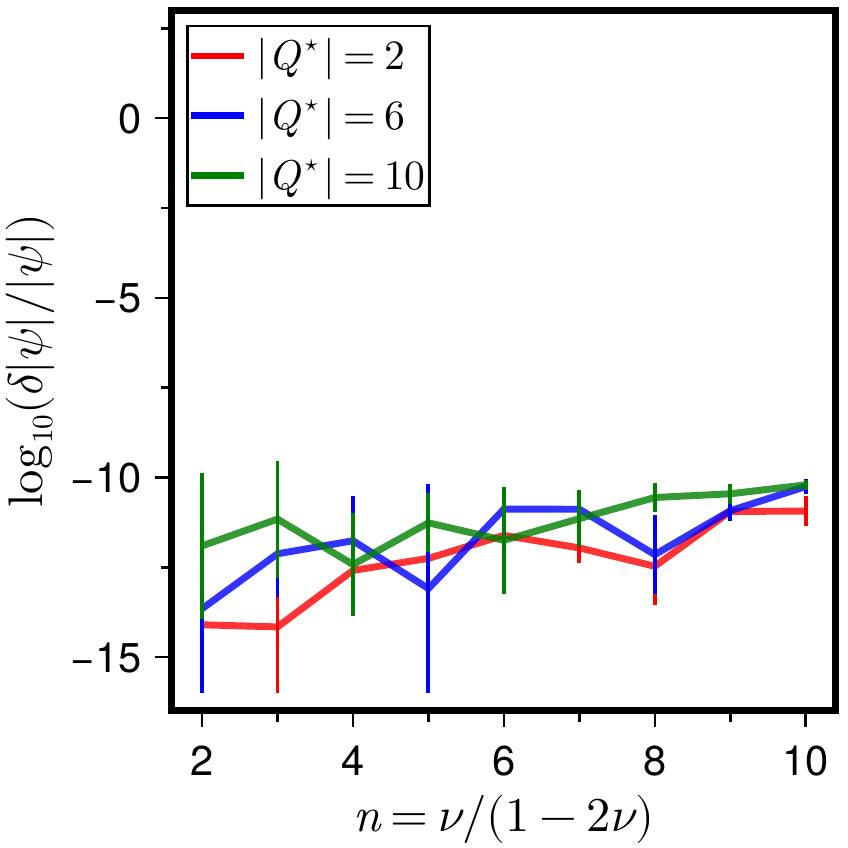}
    \caption{The top plot shows the relative error in the (absolute value of the) Jain-Kamilla (JK) projected wave function obtained using our new formulation for the IQH states of CFs as a function of $n$ [here we have $n= \nu/(1-2\nu))$, and negative values of $n$ correspond to reverse-flux attached of CFs], the number of filled $\Lambda$ levels for three different system sizes ($N = 2nQ^{\star}+n^{2}$). We have calculated the ``true'' value of the JK projected wave function by evaluating it with higher precision arithmetic. The error bars indicate the range of the relative error over $1000$ randomly generated configurations of electrons. The middle plot shows the relative error in the JK projected wave function obtained using the traditional formulation, and the bottom plot shows the same but calculated with higher precision arithmetic.}
    \label{fig:accuracy-comparison}
\end{figure}

\begin{figure}
        \includegraphics[width=0.80\columnwidth]{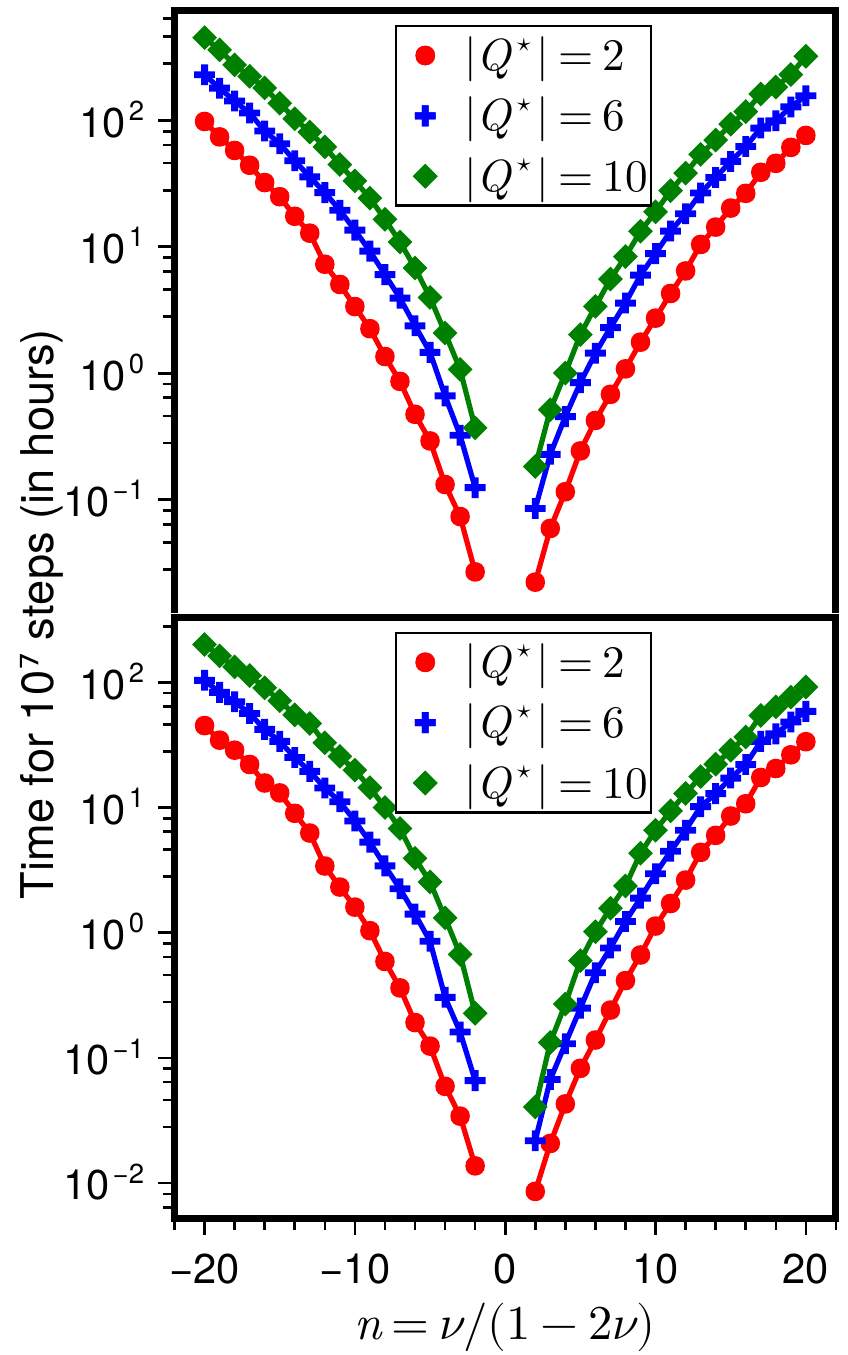}
        \caption{In this figure, we plot the time needed to run a Monte Carlo chain of length $10^{7}$ with the IQH states of CFs as the sampling wave functions as a function of $n$, the number of filled $\Lambda$ levels, for three different system sizes. We generate the Monte Carlo chain according to the Metropolis-Hastings-Gibbs algorithm. The wave functions are evaluated according to our new formulation for JK projection. In the top plot, we include the time required for a ``full" evaluation, which involves evaluation of the CF Slater matrix, the determinant of this matrix and its inverse. (The inverse is used in CF diagonalization~\cite{Mandal02} wherein the ratio between two JK projected wave functions is efficiently calculated using the matrix determinant lemma~\cite{Harville97}.) We show in the bottom plot the time required to evaluate only the CF Slater matrix.}
        \label{fig:timing-new-method}
    \end{figure}

\begin{figure}
\includegraphics[width=0.80\columnwidth]{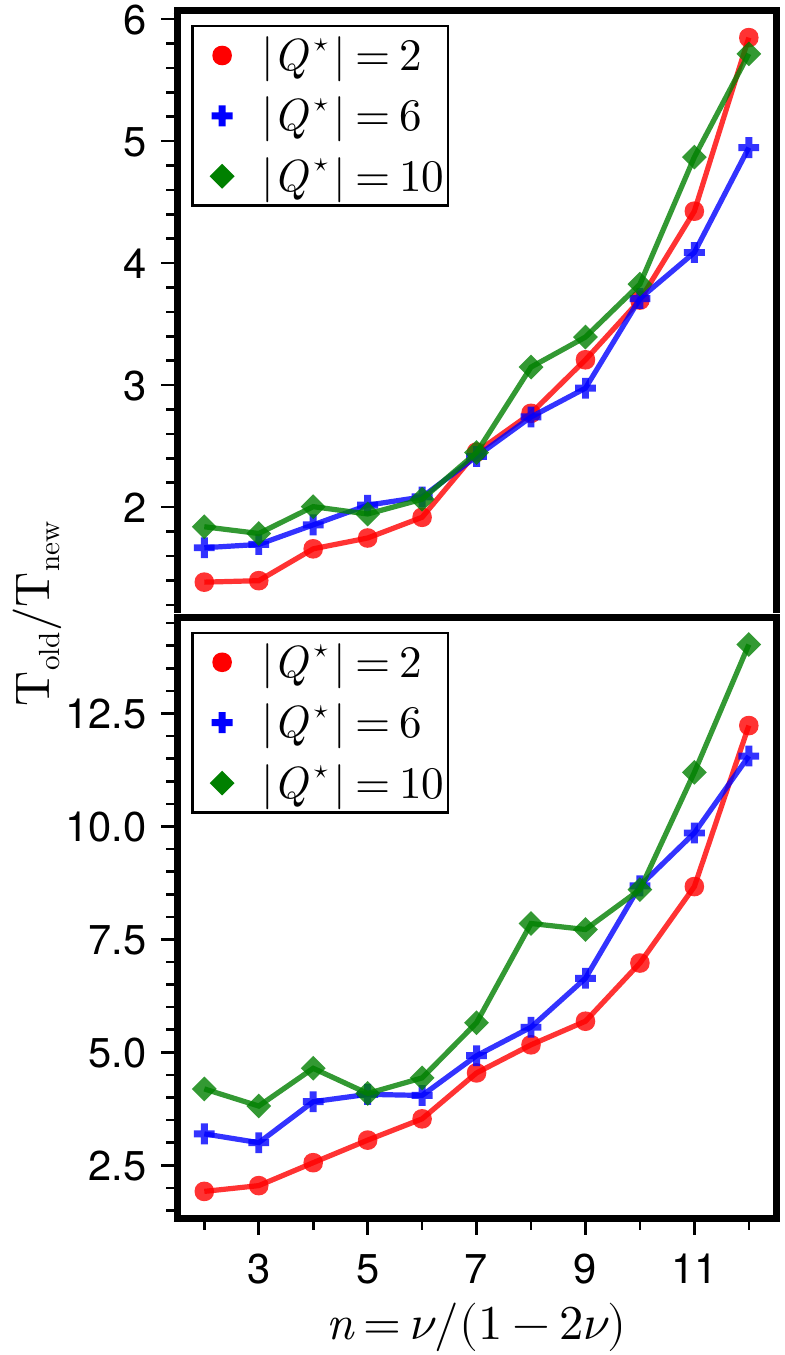}
        \caption{In this figure, we plot the ratio of the time to run a Monte Carlo chain with the IQH states of CFs as the sampling wave functions as a function of $n$, the number of filled $\Lambda$ levels (for three different system sizes) between the traditional and our new approach to JK projection. In this case, we evaluate both wave functions in the standard double precision arithmetic. The top plot, shows the ratio between times for a full evaluation and the bottom plot shows the ratio for evaluation of only the CF Slater matrix.}
        \label{fig:timing-ratios-old-to-new}
    \end{figure}

\begin{figure}
\includegraphics[width=0.80\columnwidth]{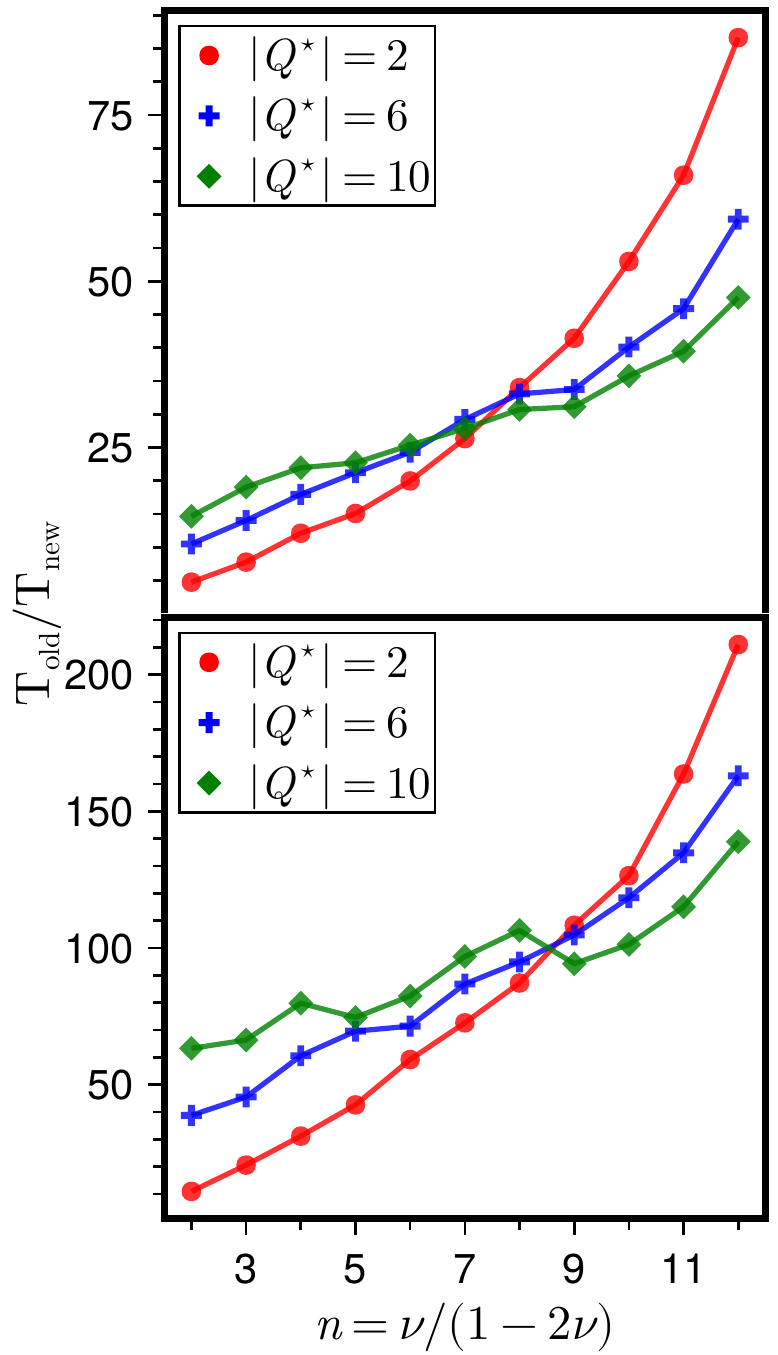}
        \caption{In this figure, we plot the ratio of the time to run a Monte Carlo chain with the IQH states of CFs as the sampling wave functions as a function of $n$, the number of filled $\Lambda$ levels between the traditional and our new approach to JK projection. In this case, we evaluate the wave function in the traditional formulation with higher precision arithmetic. The top plot, shows the ratio between times for a full evaluation and the bottom plot shows the ratio for evaluation of only the CF Slater matrix.}
        \label{fig:timing-ratios-old-higher-precision-to-new}
\end{figure}

Traditionally, the LLL projection of the elements of the determinant in the JK projected wave function is carried out as follows. First note that in $Y_{Q^{\star}, l_{j}, m_{j}}(\Omega_{i})\prod_{k \neq i}(u_{i}v_{k}-u_{k}v_{i})$ only the state of the $i^{\rm th}$ electron is possibly outside the LLL. We thus need to project wave functions of the form: $Y_{Q^{\star}, l, m}(\Omega)\left(\sum_{m=-Q_{1}}^{Q_{1}}c_{m}Y_{Q_{1}, Q_{1}, m}(\Omega)\right)$ where $c_{m}$ are arbitrary complex numbers and $Q_{1} = (N-1)/2$. This can be accomplished as follows~\cite{Jain07}:
\begin{equation}
    \begin{split}
        &\LLL Y_{Q^{\star}, l, m}(\Omega) \left(\sum_{m=-Q_{1}}^{Q_{1}}c_{m}Y_{Q_{1}, Q_{1}, m}(\Omega)\right) \\
        & = \hat{Y}_{Q^{\star}, l, m}^{Q_{1}}(\Omega) \left(\sum_{m=-Q_{1}}^{Q_{1}}c_{m}Y_{Q_{1}, Q_{1}, m}(\Omega)\right).
    \end{split}
\end{equation}
Here, $\hat{Y}_{Q^{\star}, l, m}^{Q_{1}}(\Omega)$ is an operator \textit{independent} of the LLL state it acts on, and can be obtained (up to a proportionality constant) by moving the $\conj{u}$'s and $\conj{v}$'s in each term of the polynomial expression for $Y_{Q^{\star}, l, m}(\Omega)$ (given in Eq.~\ref{eq:appendix-monopole-harmonics-sphere}) to the {\it right} and then replacing them with $\partial / \partial u$ and $\partial / \partial v$ respectively. That is:
\begin{equation}\label{eq:appendix-traditional-jk}
    \begin{split}
    &\LLL Y_{Q^{\star}, l, m}(\Omega_{i})\prod_{k \neq i}(u_{i}v_{k}-u_{k}v_{i}) \propto \\
    &\sum_{s}(-1)^{s}\binom{l-m}{s}\binom{Q^{\star}+l}{l-m-s}u_{i}^{Q^{\star}+m+s}v_{i}^{l-m-s}\times \\
    &\left[\left(\frac{\partial^{s}}{\partial u_{i}^{s}}\right)\left(\frac{\partial^{l-Q^{\star}-s}}{\partial v_{i}^{l-Q^{\star}-s}}\right)\right]\prod_{k \neq i}(u_{i}v_{k}-u_{k}v_{i}).
    \end{split}
\end{equation}
The mixed derivatives $\partial^{s}_{u_{i}}\partial^{l-Q^{\star}-s}_{v_{i}}\prod_{k \neq i}(u_{i}v_{k}-u_{k}v_{i})$ in the traditional formulation are the biggest hurdle in the numerical evaluation of the JK projected wave functions.
Firstly, the number of such derivatives that need to be evaulated grows quadratically with  $l_{\mathrm{max}}-Q^{\star}$~($l_{\mathrm{max}}$ is the maximum angular momentum of an electron in $\det Y_{Q^{\star}, l_{j}, m_{j}}(\Omega_{i})$) as $(l_{\mathrm{max}}-Q^{\star}+1)(l_{\mathrm{max}}-Q^{\star}+2)/2$. This makes it impractical (see Fig.~\ref{fig:timing-new-method} and Fig.~\ref{fig:timing-ratios-old-to-new}) to study any parallel-flux attached CF states for fillings $\nu > 8/17$ (and any reverse-flux attached CF states containing more than $\sim 40$ particles). Secondly, even when the number of derivatives required is manageable, no numerically stable algorithm is known to compute them (see Ref.~\cite{Davenport12} for a discussion). This can be seen from Fig.~\ref{fig:accuracy-comparison} where we show the accuracy in the evaluation of the wave function (evaluated using the method outlined in Ref.~\cite{Jain07}) for various values of $n = l_{\mathrm{max}}-Q^{\star}$ and system sizes. We find that for $n \geq 8$, within the standard double-precision arithmetic, the loss in accuracy makes numerical studies infeasible. Although it is possible to get accurate wave functions using higher precision arithmetic~\cite{DoubleFloats} as seen in Fig.~\ref{fig:accuracy-comparison}, it leads to a dramatic decrease in speed as seen from Fig.~\ref{fig:timing-ratios-old-higher-precision-to-new} rendering it impractical. To overcome these limitations, we now propose a new formulation based on the quaternion representation of monopole harmonics; within this formulation the computation of the CF wave functions scales linearly with $l_{\mathrm{max}}-Q^{\star}$ and is numerically stable. 

We begin by extending the JK projected wave functions to the space of unit quaternions, i.e., we replace $Y_{Q^{\star}, l, m}(\Omega_{i})$ with $\mathcal{Y}_{Q^{\star}, l, m}(r_{i})$ where $r_{i} = e^{\phi_{i}\bm{k}/2}e^{\theta_{i}\bm{y}/2}e^{\psi_{i}\bm{k}/2}$ (or equivalently, $u_{i}, v_{i}$ with the symmetric and anti-symmetric projections $r_{\mathrm{S}i}, r_{\mathrm{A}i}$) in Eq.~\ref{eq:appendix-traditional-jk}:
\begin{equation}
    \det \left[\LLL \mathcal{Y}_{Q^{\star}, l_{j}, m_{j}}(r_{i})\prod_{k \neq i}(r_{\mathrm{S}i}r_{\mathrm{A}k}-r_{\mathrm{S}k}r_{\mathrm{A}i})\right].
\end{equation}
Using $r_{i}^{-1}{}_{\mathrm{S}} = \conj{r}_{i}{}_{\mathrm{S}}$ and $ r_{i}^{-1}{}_{\mathrm{A}} = -{r}_{i}{}_{\mathrm{A}}$, we note that the symmetric and anti-symmetric projections transform under quaternion multiplication as
$$
(r_{i}^{-1} \cdot r_{j}){}_{\mathrm{S}} = \conj{r}_{i}{}_{\mathrm{S}}r_{j}{}_{\mathrm{S}} + \conj{r}_{i}{}_{\mathrm{A}}r_{j}{}_{\mathrm{A}},$$ and 
$$(r_{i}^{-1} \cdot r_{j}){}_{\mathrm{A}} = {r}_{i}{}_{\mathrm{S}}r_{j}{}_{\mathrm{A}} - {r}_{i}{}_{\mathrm{A}}r_{j}{}_{\mathrm{S}}.
$$
We can define a ``quaternion displacement'' $d(r_{i}, r_{j}) = \left(r_{i}^{-1} \cdot r_{j}\right)_{\mathrm{A}} = {r}_{i}{}_{\mathrm{S}}r_{j}{}_{\mathrm{A}} - {r}_{i}{}_{\mathrm{A}}r_{j}{}_{\mathrm{S}}$. Like the planar displacement $z_{i}-z_{j}$ (here, $z_{i} = x_{i} - \imath y_{i}$) which is invariant under translations $z_{i} \rightarrow z_{i} + z$, the quaternion displacement $d(r_{i}, r_{j})$ is {invariant} under left quaternion multiplication i.e. $d(r \cdot r_{i}, r \cdot r_{j}) = d(r_{i}, r_{j}) \; \forall r$! We now leverage this invariance as follows:
\begin{align*}
    &\mathcal{Y}_{Q^{\star}, l, m}(r_{i})\prod_{j \neq i}d(r_{i}, r_{j})=\sum_{m^{\prime}=-l}^{l}\mathcal{D}^{l}_{m, m^{\prime}}(r) \times \\
    &\mathcal{Y}_{Q^{\star}, l, m^{\prime}}(r^{-1}\cdot r_{i})\prod_{j \neq i}[(r^{-1}\cdot r_{i})^{-1}\cdot (r^{-1} \cdot r_{j})]_{\mathrm{A}}.
\end{align*}
Given that this is true for all $r$, we must have:
\begin{align*}
    &\LLL \mathcal{Y}_{Q^{\star}, l, m}(r_{i})\prod_{j \neq i}d(r_{i}, r_{j}) = \sum_{m^{\prime}=-l}^{l}\mathcal{D}^{l}_{m, m^{\prime}}(r)\times \\
    & \LLL \left\{\mathcal{Y}_{Q^{\star}, l, m^{\prime}}(r^{-1}\cdot r_{i})\prod_{j \neq i}[(r^{-1}\cdot r_{i})^{-1}\cdot (r^{-1} \cdot r_{j})]_{\mathrm{A}}\right\}.
\end{align*}
Now, let $R_{i} = r^{-1}\cdot r_{i}$ and $R_{\mathrm{S}i}, R_{\mathrm{A}i}$ be the corresponding symmetric and anti-symmetric projections. It then follows that:
\begin{align*}
    & \LLL \left\{\mathcal{Y}_{Q^{\star}, l, m^{\prime}}(r^{-1}\cdot r_{i})\prod_{j \neq i}[(r^{-1}\cdot r_{i})^{-1}\cdot (r^{-1} \cdot r_{j})]_{\mathrm{A}}\right\} \\
    &= \LLL \mathcal{Y}_{Q^{\star}, l, m^{\prime}}(R_{i})\prod_{j \neq i}(R_{\mathrm{S}i}R_{\mathrm{A}j}-R_{\mathrm{S}j}R_{\mathrm{A}i})\\
    &= \left(\prod_{j \neq i}R_{\mathrm{A}j} \right) \times \\ 
    & \LLL \mathcal{Y}_{Q^{\star}, l, m^{\prime}}(R_{i}) \sum_{r=0}^{N-1}(-1)^{r}R_{\mathrm{S}i}^{(N-1)-r}R_{\mathrm{A}i}^{r}e_{r}(\left\{\frac{R_{\mathrm{S}j}}{R_{\mathrm{A}j}}\right\}_{j \neq i}).
\end{align*}
Here, $e_{r}(X)$ is the $r^{\rm th}$ elementary symmetric polynomial constructed from the elements of the set $X$. We now have:
\begin{align*}
    & \LLL \mathcal{Y}_{Q^{\star}, l, m^{\prime}}(R_{i}) \sum_{r=0}^{N-1}(-1)^{r}R_{\mathrm{S}i}^{(N-1)-r}R_{\mathrm{A}i}^{r}e_{r}(\left\{\frac{R_{\mathrm{S}j}}{R_{\mathrm{A}j}}\right\}_{j \neq i}) \\
    & = \sum_{m^{\prime \prime}=-Q_{1}}^{Q_{1}}\frac{1}{\sqrt{\frac{2Q_{1}+1}{4\pi}\binom{2Q_{1}}{Q_{1}-m^{\prime \prime}}}}e_{Q_{1}-m^{\prime \prime}}(\left\{\frac{R_{\mathrm{S}j}}{R_{\mathrm{A}j}}\right\}_{j \neq i}) \times \\
    & \LLL \mathcal{Y}_{Q^{\star}, l, m^{\prime}}(R_{i})\mathcal{Y}_{Q_{1}, Q_{1}, m^{\prime \prime}}(R_{i}).
\end{align*}
Up until now, everything has involved rotations by an arbitrary quaternion $r$. Now, {\it and this is the key step}, we make a specific choice: $r = r_{i}^{-1}$ such that $R_{i} = \mathbf{1}$. Here the monopole harmonics satisfy $\mathcal{Y}_{Q, Q, m^{\prime}}(\mathbf{1}) = \delta_{m^{\prime}, Q}\sqrt{(2Q+1)/(4\pi)}$. And thus when $R_{i} = \mathbf{1}$, $\LLL \mathcal{Y}_{Q^{\star}, l, m^{\prime}}(R_{i})\mathcal{Y}_{Q_{1}, Q_{1}, m^{\prime \prime}}(R_{i})$ is non-zero if and only if $m^{\prime \prime} + m^{\prime} = Q^{\star} + Q_{1}$. This follows by noting that we can write $\mathcal{Y}_{Q^{\star}, l, m^{\prime}}(R_{i})\mathcal{Y}_{Q_{1}, Q_{1}, m^{\prime \prime}}(R_{i})$ as $\sum_{L}c_{L}\mathcal{Y}_{Q_{1}+Q^{\star}, L, m^{\prime}+m^{\prime \prime}}(R_{i})$~\cite{Jain07}. Therefore, we have:
\begin{align*}
    & \LLL \mathcal{Y}_{Q^{\star}, l, m^{\prime}}(R_{i}) \sum_{r=0}^{N-1}(-1)^{r}R_{\mathrm{S}i}^{(N-1)-r}R_{\mathrm{A}i}^{r}e_{r}(\left\{\frac{R_{\mathrm{S}j}}{R_{\mathrm{A}j}}\right\}_{j \neq i}) \\
    & = \frac{1}{\sqrt{\frac{2Q_{1}+1}{4\pi}\binom{2Q_{1}}{m^{\prime}-Q^{\star}}}}e_{m^{\prime}-Q^{\star}}(\left\{\frac{(r_{i}^{-1}\cdot r_{j})_{\mathrm{S}}}{(r_{i}^{-1}\cdot r_{j})_{\mathrm{A}}}\right\}_{j \neq i}) \times \\
    & \left[\LLL \mathcal{Y}_{Q^{\star}, l, m^{\prime}}(r)\mathcal{Y}_{Q_{1}, Q_{1}, Q^{\star}+Q_{1}-m^{\prime}}(r)\right]_{r \rightarrow \mathbf{1}}\;.
\end{align*}
Let us note that~\cite{Jain07}
\begin{align*}
    &\LLL \mathcal{Y}_{Q^{\star}, l, m^{\prime}}(r)\mathcal{Y}_{Q_{1}, Q_{1}, Q^{\star}+Q_{1}-m^{\prime}}(r)\\
    &=(-1)^{l+2Q_{1}+Q^{\star}}\\
    &\times \sqrt{\frac{(2l+1)(2Q_{1}+1)(2Q_{1}+2Q^{\star}+1)}{4\pi}} \\
    &\times \begin{pmatrix}
    l & Q_{1} & Q_{1} + Q^{\star} \\
    -m^{\prime} & m^{\prime}-Q_{1}-Q^{\star} & Q_{1} + Q^{\star}
    \end{pmatrix}\\
    &\times \begin{pmatrix}
    l & Q_{1} & Q_{1} + Q^{\star} \\
    Q^{\star} & Q_{1} & -Q^{\star}-Q_{1}
    \end{pmatrix}\mathcal{Y}_{Q^{\star}+Q_{1}, Q^{\star}+Q_{1}, Q^{\star}+Q_{1}}(r).
\end{align*}
Here, the $2 \times 3$ matrices correspond to the $3j$ symbols, which can be calculated conveniently via a symbolic manipulation software to obtain:
\begin{align*}
    &\frac{\left(\LLL \mathcal{Y}_{Q^{\star}, l, m^{\prime}}(r)\mathcal{Y}_{Q_{1}, Q_{1}, Q^{\star}+Q_{1}-m^{\prime}}(r)\right)_{r \rightarrow \mathbf{1}}}{\sqrt{\frac{2Q_{1}+1}{4\pi}\binom{2Q_{1}}{m^{\prime}-Q^{\star}}}}\\
    &= (-1)^{m^{\prime}-Q^{\star}}\sqrt{\frac{2l+1}{4\pi}}\frac{(2Q_{1} + 2Q^{\star}+1)!}{(2Q_{1}+Q^{\star}+l+1)!}\\
    &\times \frac{(2Q_{1}+Q^{\star}-m^{\prime})!}{(2Q_{1}+Q^{\star}-l)!}\sqrt{\frac{(l-Q^{\star})!}{(l-m^{\prime})!}\frac{(l+m^{\prime})!}{(l+Q^{\star})!}}.
\end{align*}
We therefore find that the elements of the Slater determinant in the JK wave function (see Eq.~\ref{eq:appendix-traditional-jk}) are equal to:
\begin{equation}\label{eq:appendix-jk-new}
\begin{split}
    &{\left[\LLL \mathcal{Y}_{Q^{\star}, l, m}(r_{i})\prod_{j \neq i}d(r_{i}, r_{j})\right]}/{\prod_{j \neq i}d(r_{i}, r_{j})}\\
    &= \sum_{m^{\prime}=Q^{\star}}^{l}(-1)^{m^{\prime}-Q^{\star}}N^{l}_{m^{\prime}, Q^{\star}, Q_{1}}\mathcal{D}_{m, m^{\prime}}^{l}(r_{i})\tilde{e}_{m^{\prime}-Q^{\star}}(X_{i})\\
    &N^{l}_{m^{\prime}, Q^{\star}, Q_{1}} \\
    &= \left({\binom{2Q_{1}}{l-Q^{\star}}\binom{l-Q^{\star}}{m^{\prime}-Q^{\star}}}/{\binom{2Q_{1}+l+Q^{\star}+1}{l-Q^{\star}}}\right)\\
    &\times \sqrt{{\binom{2l}{l+Q^{\star}}}/{\binom{2l}{l+m^{\prime}}}} \times \sqrt{(2l+1)/(4\pi)}.
\end{split}
\end{equation}
Here, $\tilde{e}_{r}(X_{i}) \equiv e_{r}(X_{i})/(\binom{N-1}{r})$ are a regularized version of the elementary symmetric polynomials $e_{m^{\prime}-Q^{\star}}(X_{i})$. $X_{i}$ is the set $\left\{(r_{i}^{-1}\cdot r_{j})_{\mathrm{S}}/(r_{i}^{-1}\cdot r_{j})_{\mathrm{A}} \vert j \neq i, j = 1, \dots, N\right\}$. Importantly, if $l_{\mathrm{max}}$ is the maximum angular momentum of an electron in $\det[Y_{Q,l_{j},m_{j}}(r_{i})]$, we need to evaluate only $l_{\mathrm{max}}-Q^{\star}+1$ elementary symmetric polynomials (for each column)! The elementary symmetric polynomials themselves can be computed very efficiently by a dynamic programming algorithm~\cite{Davenport12, Macdonald79} that leverages the following relationship: $e_{k}(\left\{x_{1}, x_{2}, \dots, x_{N}\right\}) = e_{k-1}(\left\{x_{2}, \dots, x_{N}\right\}) + x_{1} e_{k-1}(\left\{x_{2}, \dots, x_{N}\right\})$. Given that the elements of the set $X_{i}$, for example, equal to $\left\{(\conj{u}_{i}u_{j} + \conj{v}_{i}v_{j})/(u_{i}v_{j}-u_{j}v_{i}) \vert j \neq i\right\}$ on the $\psi_{i} = 0$ sub-space, are well-behaved (during a Monte Carlo run with the sampling distribution as the wavefunction) because of the Pauli principle, it follows that the (regularized) elementary symmetric polynomials appearing in Eq.~\ref{eq:appendix-jk-new} can always be safely computed within standard double-precision arithmetic. Further, from the Fourier decomposition of Wigner-$d$ matrices proposed in Ref.~\cite{Feng15},
\begin{align*}
    & d^{l}_{m, m^{\prime}}(\theta) = \bra{l, m}e^{-\imath \theta J_{y}}\ket{l, m^{\prime}} \\
    & = \bra{l, m}\left(\sum_{\mu^{\prime}}{}_{y}\ket{l,\mu^{\prime}}\bra{l, \mu^{\prime}}_{y}\right)e^{-\imath \theta J_{y}}\\
    &\times \left(\sum_{\mu}{}_{y}\ket{l, \mu}\bra{l, \mu^{\prime}}_{y}\right)\ket{l, m^{\prime}} \\
    & = \sum_{\mu = -l}^{l}e^{-\imath \mu \theta}\overlap{l,m}{l,\mu}_{y}{}_{y}\overlap{l,\mu}{l,m^{\prime}}
\end{align*}
where $\ket{l, \mu}_{y}$ are the eigenstates of $J_{y}$ (obtained through exact diagonalization of $J_{y}$ in the $J_{z}$ eigenstate basis), we can calculate the Wigner-$\mathcal{D}$ matrices appearing in Eq.~\ref{eq:appendix-jk-new} to an accuracy of $10^{-14}$ (for $l_{\mathrm{max}} \leq 100$). The accuracy in calculating the elementary symmetric polynomials and the Wigner-$\mathcal{D}$ matrices, alongside the linear scaling in $l_{\mathrm{max}}-Q^{\star}$ leads to a dramatic increase in both the speed and accuracy of the evaluation of the JK projected wave functions as seen from Fig.~\ref{fig:accuracy-comparison}, Fig.~\ref{fig:timing-new-method}, Fig.~\ref{fig:timing-ratios-old-to-new} and Fig.~\ref{fig:timing-ratios-old-higher-precision-to-new}. This allows us to study for the first time any Jain FQH state at filling $\nu$ in the range $10/21, \dots, 20/41$ or $20/39, \dots, 6/11$. Further, even for fillings previously accessible, our new method allows us to access much larger system sizes than before, paving the way for studying the thermodynamic behavior of observables with complicated or unknown system size dependence.

\section{Magneto-roton modes of Jain FQH states}\label{appendix-roton-dispersion}

In this section, we briefly review the Jain $\nu = n/(2pn + 1)$ states and their lowest energy charged and neutral excitations (the latter are also referred to as the magneto-roton modes) in the spherical geometry. To explore the possibility of a magneto-roton instability of the FQHE along the sequence $\nu = n/(2n+1)$ for large $n$, we have evaluated the excitation energies of the magneto-roton modes for fillings $\nu = 1/3, 2/5, \dots, 10/21$ and $\nu = 15/31$ using our new formulation. These results show no evidence of a magneto-roton instability.

Electrons constrained to move within the LLL (for filling factors $\nu \leq 1$) capture $2p$ (here, $p \in \mathbb{Z}^{+}$) vortices to form CFs. As a result of the vortex capture, CFs experience a reduced magnetic field $B^{*}=B-2p\rho \phi_{0}$. Here, $B$ is the applied magnetic field, and $\rho$ is the electron density. CFs experiencing the reduced field $B^{*}$ inhabit Landau-like levels called $\Lambda$ levels called $\Lambda$ levels ($\Lambda$Ls) that are separated by a nearly constant energy difference called the CF cyclotron energy $\hbar \omega_{c}^{*} \propto B^{\star}$. When CFs fill the first $n$ CF LLs (called $\Lambda$Ls), corresponding to electrons at filling $\nu = n/(2pn+1)$, the system becomes incompressible. (Note, that by definition the filling factor $\nu^{\star}$ of CFs is related to the electronic filling as $\nu^{\star} = \nu / (1-2p\nu) \Leftrightarrow \nu = \nu^{\star}/(1+2p\nu^{\star})$.) That is, the integer quantum Hall effect (IQHE) of CFs results in the FQHE of electrons. An incompressible state of $N$ CFs at monopole strength $Q^{\star} = N/2n-n/2$ filling $n$ $\Lambda$Ls (i.e. $\nu^{\star}=n$) can be written as
    \begin{equation}\label{eq:cf-ground-state}
        \begin{split}
            &\Psi_{n/(2pn+1)}(\Omega_{1}, \dots, \Omega_{N}) \\
            &= \LLL \phi_{n}(\Omega_{1}, \dots, \Omega_{N}) \times \left[ \phi_{1}(\Omega_{1}, \dots, \Omega_{N}) \right]^{2p} \\
            &\phi_{1}(\Omega_{1}, \dots, \Omega_{N}) = \prod_{i<j=1}^{N}(u_{i}v_{j}-u_{j}v_{i}).
        \end{split}    
    \end{equation}
Here, $\Omega_{i} \equiv (\theta_{i}, \phi_{i})$ is the position of the $i^{\rm th}$ electron on the sphere and $u_{i}, v_{i}$ the corresponding spinor variables. $\LLL$ is the projection operator into the LLL. $\phi_{n}(\Omega_{1}, \dots, \Omega_{N})$ is a Slater determinant of $N$ electrons filling the first $n$ Landau levels at monopole strength $Q^{\star}$ and $\phi_{1}^{2p}$ is the Jastrow factor which attaches $2p$ vortices to each electron and transforms them into CFs.

The lowest energy negatively charged excitation called the quasiparticle at filling $\nu = n/(2pn+1)$ is obtained by adding a CF to the lowest unoccupied (i.e. the $n^{\rm th}$) $\Lambda$L, and a positively charged excitation called the quasihole is obtained by  removing a CF from the highest occupied (i.e. the $(n-1)^{\rm th}$) $\Lambda$L. A state containing a CF added to the $n^{\rm th}$ $\Lambda$L in the $L_{z}=m$ orbital is written as
    \begin{equation}\label{eq:cf-qp-state}
        \begin{split}
            &\Psi^{\mathrm{qp}, m}_{n/(2pn+1)}(\Omega_{1}, \dots, \Omega_{N})\\
            &= \LLL \phi^{\mathrm{qp}, m}_{n}(\Omega_{1}, \dots, \Omega_{N}) \times \left[ \phi_{1}(\Omega_{1}, \dots, \Omega_{N}) \right]^{2p}.
        \end{split}    
    \end{equation}
Here, $Q^{\star} = (N-1)/2n-n/2$ and $\phi_{n}^{\mathrm{qp}, m}$ is a Slater determinant of $N$ electrons filling the first $n$ Landau levels at monopole strength $Q^{\star}$ and the $L_{z}=m$ orbital in the $n^{\rm th}$ Landau level. Similarly, a state with a CF removed from the $(n-1)^{\rm th}$ $\Lambda$L in the $L_{z}=m$ orbital is written as
    \begin{equation}\label{eq:cf-qh-state}
        \begin{split}
            &\Psi^{\mathrm{qh}, m}_{n/(2pn+1)}(\Omega_{1}, \dots, \Omega_{N})\\
            &= \LLL \phi^{\mathrm{qh}, m}_{n}(\Omega_{1}, \dots, \Omega_{N}) \times \left[ \phi_{1}(\Omega_{1}, \dots, \Omega_{N}) \right]^{2p}.
        \end{split}    
    \end{equation}
Here, $Q^{\star} = (N+1)/2n-n/2$ and $\phi_{n}^{\mathrm{qh}, m}$ is a Slater determinant of $N$ electrons filling the first $n$ Landau levels at monopole strength $Q^{\star}$ except for the $L_{z}=m$ orbital in the $(n-1)^{\rm th}$ Landau level. A CF exciton state containing a CF quasihole in the $(n-1)^{\rm th}$ $\Lambda$L in the $L_{z}=m$ orbital and a CF quasiparticle in $n^{\rm th}$ $\Lambda$L in the $L_{z}=m^{\prime}$ orbital is written as
    \begin{equation}\label{eq:cf-exciton-state}
        \begin{split}
            &\Psi^{\mathrm{ex}, m, m^{\prime}}_{n/(2pn+1)}(\Omega_{1}, \dots, \Omega_{N})\\
            &= \LLL \phi^{\mathrm{ex},m,m^{\prime}}_{n}(\Omega_{1}, \dots, \Omega_{N}) \times \left[ \phi_{1}(\Omega_{1}, \dots, \Omega_{N}) \right]^{2p}.
        \end{split}    
    \end{equation}
Here, $\phi^{\mathrm{ex},m,m^{\prime}}_{n}$ is a Slater determinant containing $N$ electrons at monopole strength $Q^{\star}$ filling the first $n$ Landau levels and the $L_{z}=m^{\prime}$ orbital in the $(n+1)^{\rm th}$ $\Lambda$L, with an electron missing from the $L_{z}=m$ orbital in the $(n-1)^{\rm th}$ $\Lambda$L.

The lowest energy neutral excitations i.e. the magneto-roton modes are the angular momentum eigenstates constructed from the (lowest energy) CF exciton states as:
\begin{equation}
\begin{split}
    &\Psi^{\mathrm{ex}, L,m}_{n/(2pn+1)}(\Omega_{1}, \Omega_{2}, \dots, \Omega_{N}) \\
    &= \LLL \left(\sum_{m_{1}=-l_{\mathrm{qh}}}^{m=l_{\mathrm{qh}}}\overlap{L,m}{l_{\mathrm{qh}},-m_{1};l_{\mathrm{qp}},m_{1}+m}\right. \\
    &\left. \times \phi^{\mathrm{ex},m_{1}+m,-m_{1}}_{n}(\Omega_{1}, \dots, \Omega_{N})\right)\left[\phi_{1}(\Omega_{1}, \dots, \Omega_{N})\right]^{2p}.
\end{split}
\end{equation}
Here, $L$ (related to the planar momentum $k$ as $k = L/\sqrt{Q}$) is the angular momentum of the magneto-roton, $m$ is its azimuthal quantum number. $l_{\mathrm{qh}} = Q^{\star}+n-1$ and $l_{\mathrm{qp}}=Q^{\star}+n$ are the angular momentum of the CF QH and QP respectively. $\overlap{L,0}{l_{\mathrm{qh}},-m;l_{\mathrm{qp}},m}$ are the Clebsch-Gordon coefficients. 

If the CFs were completely non-interacting, the magneto-roton mode would be dispersionless with an excitation energy equal to $\hbar \omega_{c}^{\star}$ (the kinetic energy of CFs) relative to the CF integer quantum Hall (IQH) state (see Eq.~\ref{eq:cf-ground-state}). However, in reality, the CFs do experience a residual interaction~\cite{Balram13}, which results in a dispersion of the magneto-roton mode. Along the sequence $\nu = n/(2n+1)$, as the effective magnetic field felt by CFs decreases (recall $Q^{\star} \approx Q/(2n+1)$), the excitation energy of the magneto-roton mode also decreases. This raises an important question: Could the residual interaction between CFs drive a magneto-roton instability (i.e., result in a magneto-roton with negative excitation energy) for sufficiently large $n$? An explicit calculation of the magneto-roton dispersion is necessary to answer this, as such an outcome cannot be ruled out a priori.

\begin{figure*}
\begin{tabular}{ccc}
     \includegraphics[width=0.3\textwidth]{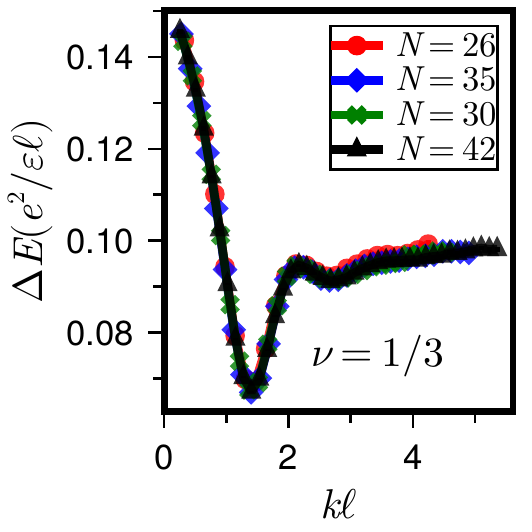}
&  \includegraphics[width=0.3\textwidth]{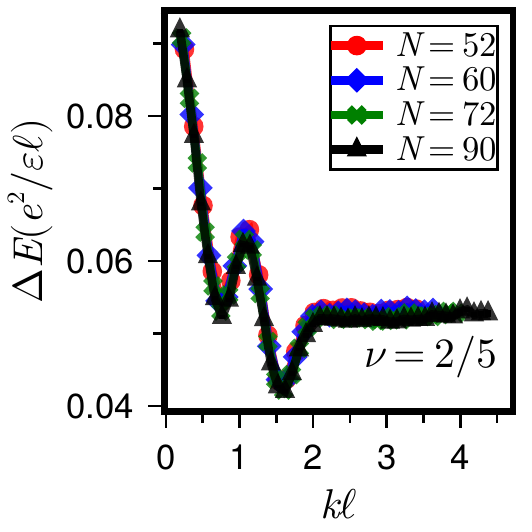}
&  \includegraphics[width=0.3\textwidth]{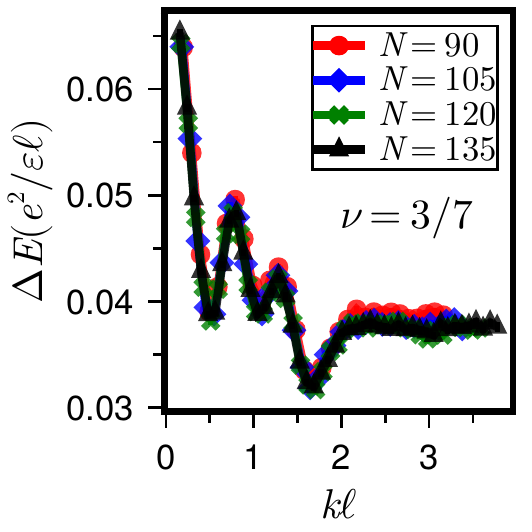} \\
     \includegraphics[width=0.3\textwidth]{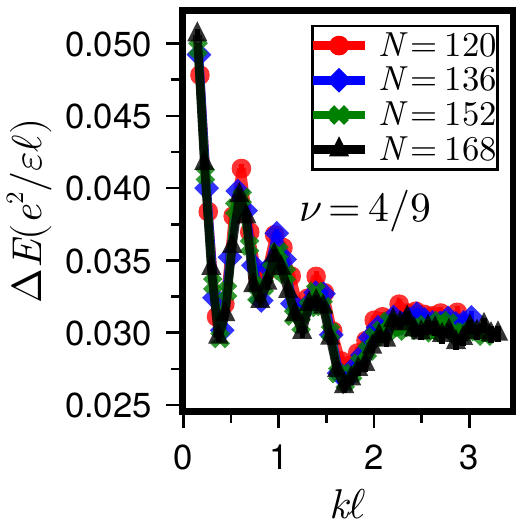}
&  \includegraphics[width=0.3\textwidth]{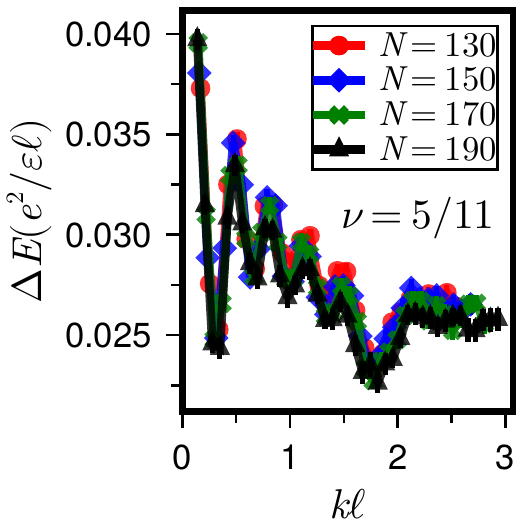}
&  \includegraphics[width=0.3\textwidth]{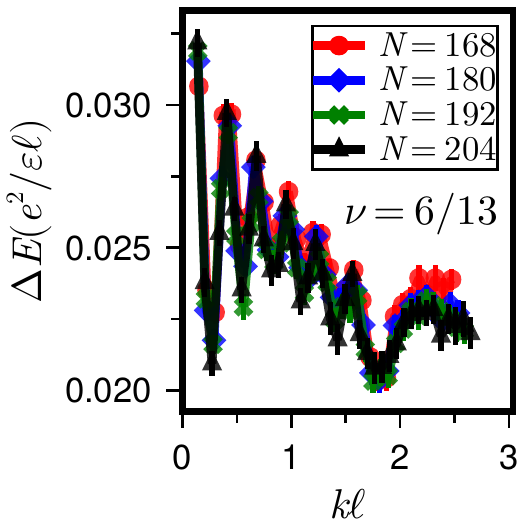} \\
     \includegraphics[width=0.3\textwidth]{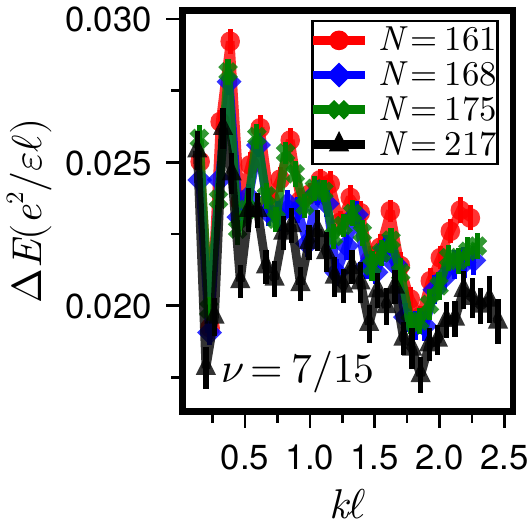}
&  \includegraphics[width=0.3\textwidth]{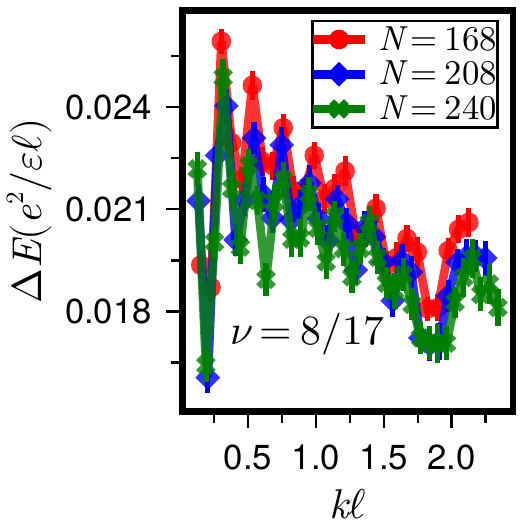}
&  \includegraphics[width=0.3\textwidth]{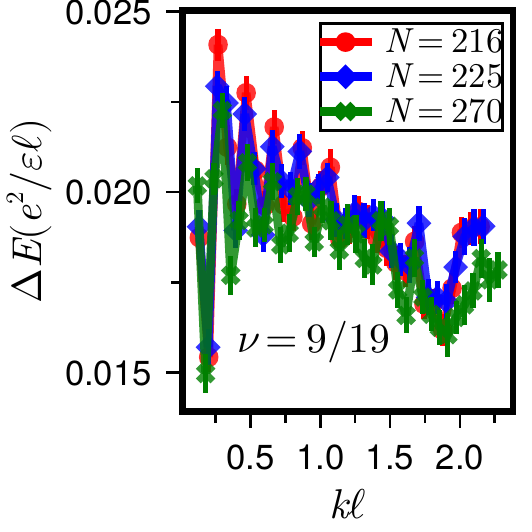} \\
\includegraphics[width=0.3\textwidth]{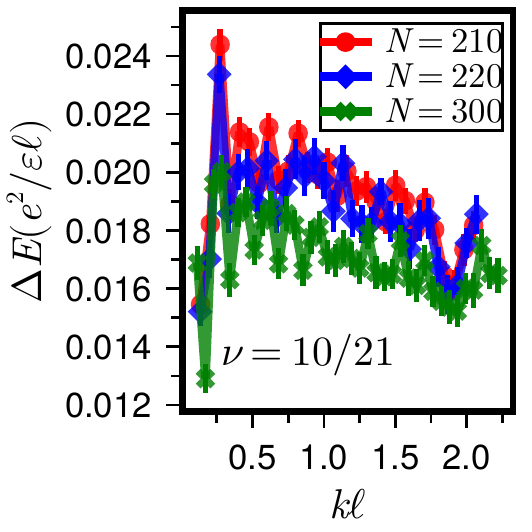} 
& \includegraphics[width=0.3\textwidth]{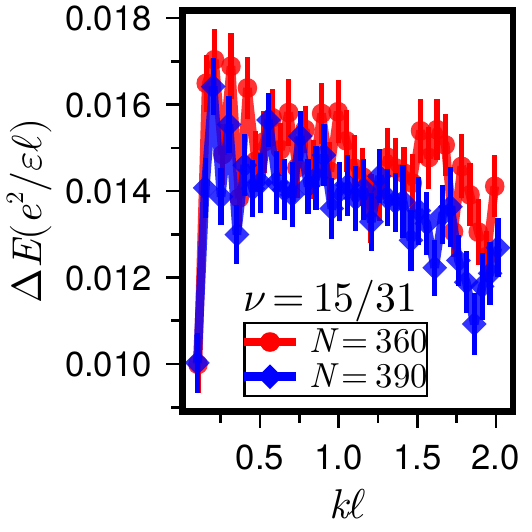} 
\end{tabular}
\caption{This figure shows the excitation energy $\Delta E$ of the magneto-roton modes at $\nu = 1/3, 2/5, \dots, 10/21$ and $\nu=15/31$ as a function of their momentum $k\ell$ for systems with $N$ electrons. Vertical bars indicate Monte Carlo uncertainty.}
\label{fig:finite-size-dispersion}
\end{figure*}

For the interaction $\hat{V}_{I}=\sum_{i<j=1}^{N}V_{I}(|r_{i}-r_{j}|)$ with $V_{I}=1/|r_{i}-r_{j}|$, we calculate the energy (recall that after LLL projection the kinetic energy of the electrons is necessarily quenched) of a wave function $\Psi$ (either equal to $\Psi_{n/(2n+1)}$ or $\Psi^{\mathrm{ex}, L, 0}_{n/(2n+1)}$) as $\sum\left[ \abs{\Psi / \Psi_{n/(2n+1)}}^{2} \hat{V}_{I}\right] / \sum \left[\abs{\Psi / \Psi_{n/(2n+1)}}^{2}\right]$. Here, $\sum$ is a sum over electronic configurations generated by sampling with respect to $\abs{\Psi_{n/(2n+1)}}^{2}$ according to the Metropolis-Hastings-Gibbs algorithm following our new method for JK projection. We generate $\approx 1.5 \times 10^{7}$ samples across $100$ Monte Carlo chains to calculate the excitation energy $\Delta E$ (i.e. the energy wrt to $\Psi_{n/(2n+1)}$) of magneto-roton modes in Fig.~\ref{fig:finite-size-dispersion} and estimate the error using the Jackknife resampling technique~\cite{Ramachandran21}. To eliminate some of the system-size dependence in our results, we account for the density correction~\cite{Morf86} i.e. place the electrons on a sphere of radius $\sqrt{N/(2\nu)}\ell$ in the calculation of $\Delta E$.

As can be seen from Fig.~\ref{fig:finite-size-dispersion}, the magneto-roton dispersions at Jain fillings $\nu = n/(2n+1)$ reveal that as we move from $\nu = 1/3$ to $\nu = 1/2$, there is a competition between the minima at $k\ell \rightarrow 0$ and $k \ell \rightarrow 2$ towards gap closing. For $\nu \leq 7/15$, the lowest energy excitation occurs at the minimum at $k\ell \sim 2$, but for $\nu \geq 8/17$, the lowest energy excitation occurs at $k\ell \sim 0$. Importantly, we find no evidence of magneto-roton instability for all the systems we have studied. 

\pagebreak

\bibliographystyle{unsrtnat}
\bibliography{biblio_fqhe.bib}